\documentclass[reprint,twocolumn,showpacs,preprintnumbers,amsmath, aps,pre,amssymb]{revtex4-1}
\usepackage{graphicx}
\usepackage{epstopdf}
\usepackage{subfigure}
\usepackage{amsmath}
\usepackage{algorithm}
\usepackage{algorithmic}
\usepackage{subfigure}
\usepackage{booktabs}
\usepackage{topcapt}
\usepackage{comment} 

\newcommand{\avg}[1]{\left\langle #1 \right\rangle}
\newcommand\kpr{k^{\prime}}
\newcommand{\xpr}{x^{\prime}}

\newcommand\bcmdtab{\noindent\bgroup\tabcolsep=0pt%
  \begin{tabular}{@{}p{10pc}@{}p{20pc}@{}}}
\newcommand\ecmdtab{\end{tabular}\egroup}

 \newcommand{\remove}[1]{}
\usepackage{color}

\newcommand{\sectA}[1]{Appendix Sec.~\ref{#1}}

\newcommand{\figwidth}{1.9in}

\usepackage{color}

\begin{document}

\title{The Majority Illusion in Social Networks}


\author{Kristina Lerman}
\author{Xiaoran Yan}
\author{Xin-Zeng Wu }
\affiliation{%
USC Information Sciences Institute, 
4676 Admiralty Way, Marina del Rey, CA 90292
}%

\date{\today}

\label{firstpage}

\begin{abstract}

Social behaviors are often contagious, spreading through a population as individuals imitate the decisions and choices of others. A variety of global phenomena, from innovation adoption to the emergence of social norms and political movements, arise as a result of people following a simple local rule, such as copy what others are doing. However, individuals often lack global knowledge of the behaviors of others and must estimate them from the observations of their friends' behaviors. In some cases, the structure of the underlying social network can dramatically skew an individual's local observations, making a behavior appear far more common locally than it is globally. We trace the origins of this phenomenon, which we call ``the majority illusion,'' to the friendship paradox in social networks. As a result of this paradox, a behavior that is globally rare may be systematically overrepresented in the local neighborhoods of many people, i.e., among their friends. Thus, the ``majority illusion'' may facilitate the spread of social contagions in networks and also explain why systematic biases in social perceptions, for example, of risky behavior, arise. Using synthetic and real-world networks, we explore how the ``majority illusion'' depends on network structure and develop a statistical model to calculate its magnitude in a network.
\end{abstract}
\maketitle

%

\section{Introduction}
An individual's behavior often depends on the actions of others~\cite{Schelling73,Granovetter78,Salganik06,Centola10,Young11,Centola15}. This phenomenon is manifested daily in the decisions people make to adopt a new technology~\cite{Valente95book,Rogers03} or idea~\cite{Bettencourt06,Young11}, listen to music~\cite{Salganik06}, engage in risky behavior~\cite{Bearak14}, or join a social movement~\cite{Granovetter78,Schelling73}. As a result, a variety of behaviors are said to be `contagious', because they spread through the population as people observe others adopting the behavior and then adopt it themselves. In some cases, behaviors will spread from a small number of initial adopters to a large portion of the population, resulting in fads, hit songs, successful political campaigns, epidemics, and social norms. Researchers have examined the conditions under which such global outbreaks occur, especially in a networked setting, where individuals interact with a subset of
the population, i.e., their network neighbors or friends. Studies have linked the onset of global outbreaks to the topology of underlying network~\cite{Watts02,Centola15}, including the presence of highly connected individuals~\cite{Kempe03,Lloyd-Smith05} and small clusters of connected people~\cite{Centola10,Young11}.
However, network structure can affect the emergence of global outbreaks in a subtler way. As we show in this paper, the configuration of initial adopters 
on a network can systematically skew the observations people make of their friends' behavior. This can
make some behavior appear much more popular than it is, thus creating conditions for its spread.

Networks often have counter-intuitive properties. One of the better known of these is the friendship paradox, which states that, on average, most people have fewer friends than their friends have~\cite{Feld91}. Despite its almost nonsensical nature, the friendship paradox has been used to design efficient strategies for vaccination~\cite{Cohen03}, social intervention~\cite{Kim15}, and early detection of contagious outbreaks~\cite{Christakis10,GarciaHerranz14}. In a nutshell, rather than monitoring random people to catch a contagious outbreak in its early stages, the friendship paradox suggests monitoring their random friends, because they are more likely to be better connected and not only to get sick earlier, but also to infect more people once sick.

Recently, friendship paradox was generalized for attributes other than degree, i.e., number of network neighbors. For example, your co-authors are cited more often than you~\cite{Eom14}, and the people you follow on Twitter post more frequently than you do~\cite{Hodas13icwsm}. In fact, any attribute that is correlated with degree will
produce a paradox~\cite{Eom14,Kooti14icwsm}. Thus, if heavy drinkers also happen to be more popular, then people examining their friends' drinking behavior will conclude that, on average, their friends drink more than they do. This may help explain why adolescents systematically overestimate their friends' alcohol consumption and drug use~\cite{Baer91,berkowitz2005overview}.

In this paper, we describe a novel variation of the friendship paradox that is essential for understanding contagious behaviors. The paradox applies to networks in which nodes have attributes, in the simplest case a binary attribute, such as ``has red hair'' vs ``does not have red hair'' or ``purchased an iPhone'' vs ``did not purchase an iPhone''. We refer to nodes that have this attribute as ``active'', and the rest are ``inactive.'' We show that under some conditions, a large fraction of nodes will observe most of their neighbors in the active state, even when it is globally rare. For this reason, we call the paradox the ``majority illusion.''

\begin{figure*}[htbp] 
\centering
\subfigure[\label{fig:net14cor}]{\includegraphics[width=\figwidth]{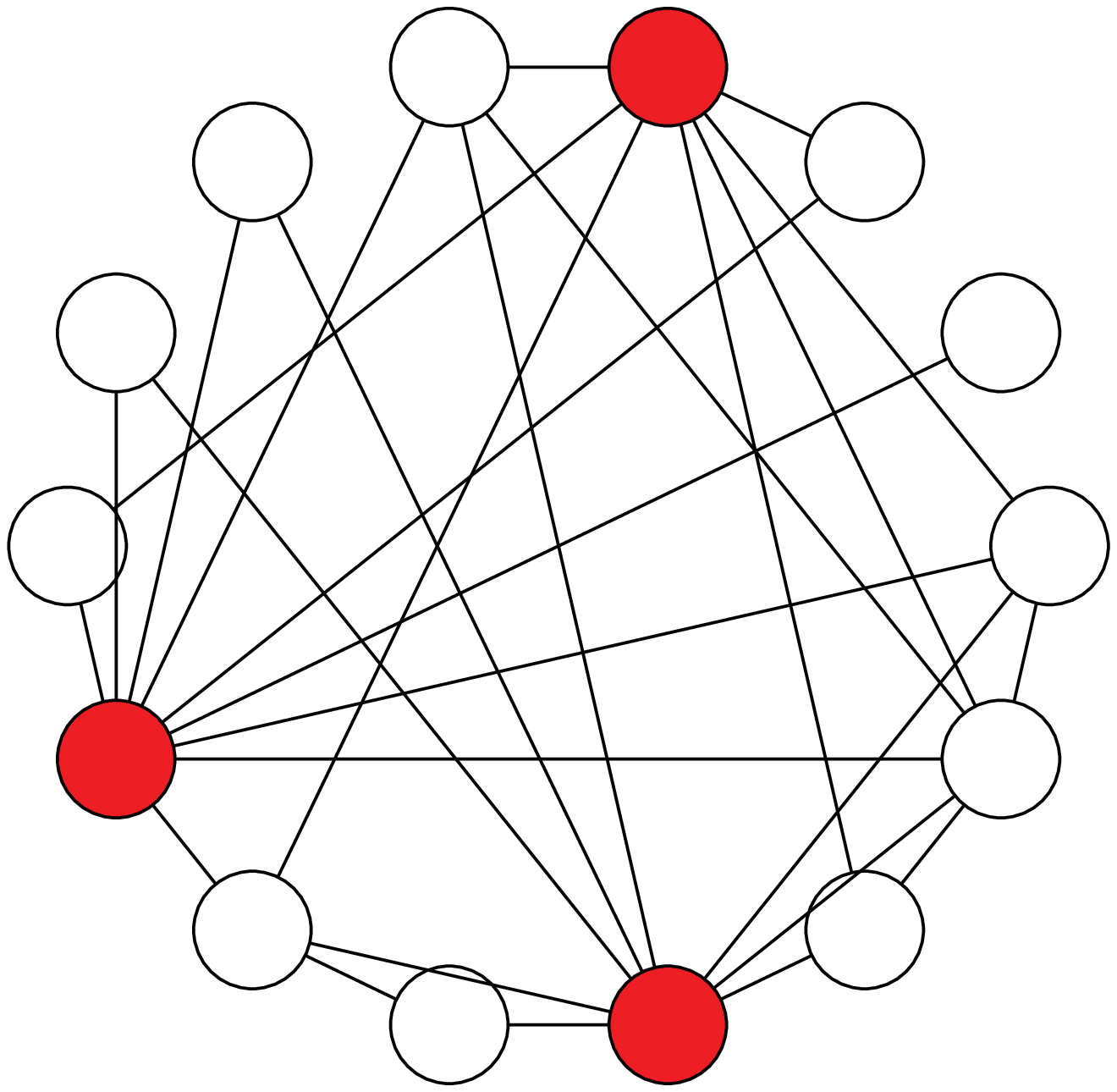}}
\subfigure[\label{fig:net14uncor}]{\includegraphics[width=\figwidth]{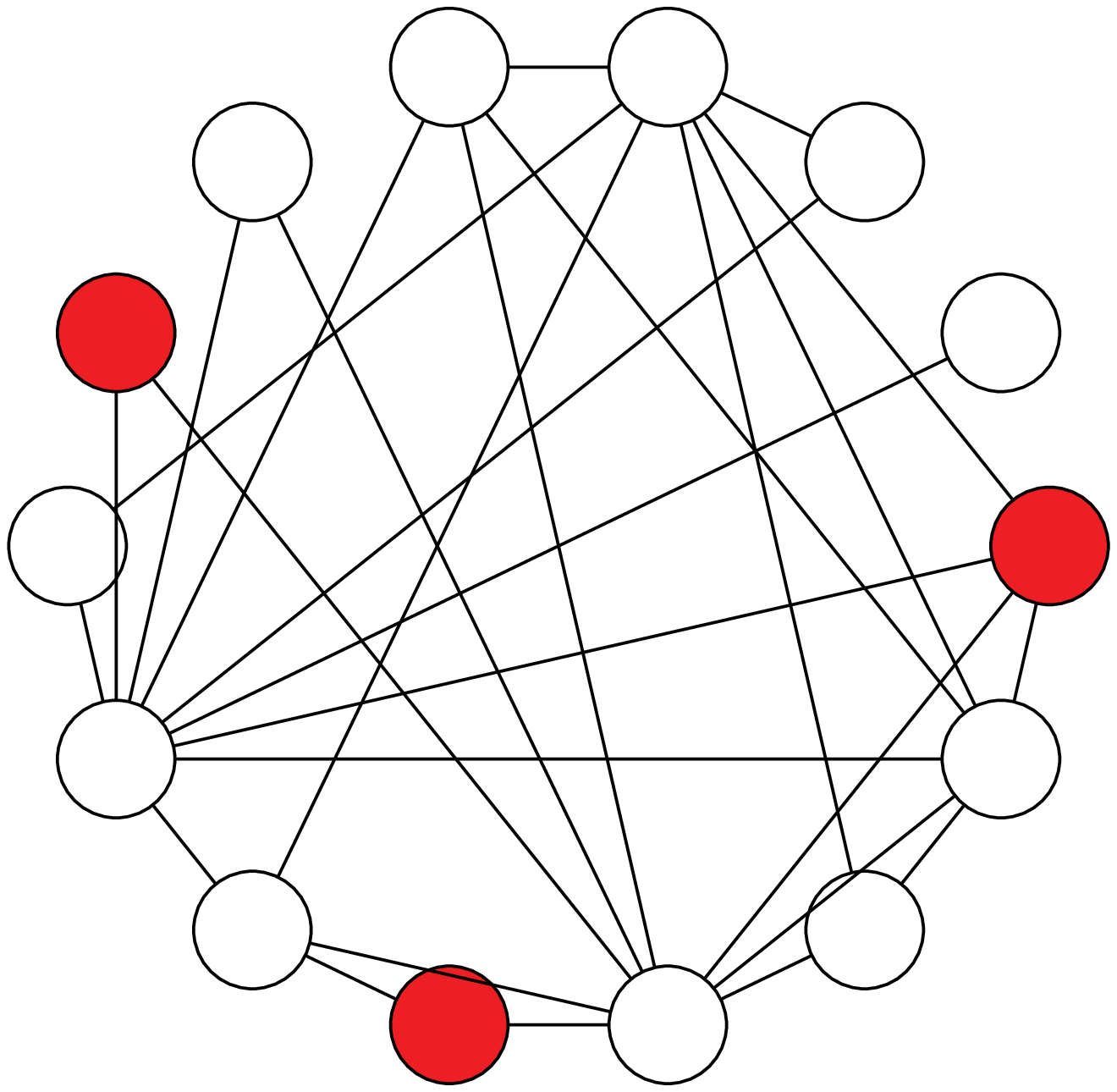}}
    \caption{An illustration of the ``majority illusion'' paradox. The two networks are identical, except for which three nodes are colored. These are the ``active'' nodes and the rest are ``inactive.'' In the network on the left, all ``inactive'' nodes observe that at least half of their neighbors are ``active,'' while in the network on the right, no ``inactive'' node makes this observation.}
    \label{fig:net14}
 \end{figure*}

As a simple illustration of the ``majority illusion'' paradox, consider the two networks in Figure~\ref{fig:net14}. The networks are identical, except for which of the few nodes are colored. Imagine that colored nodes are active and the rest of the nodes are inactive. Despite this apparently small difference, the two networks are profoundly different: in the first network, every inactive node will examine its neighbors to observe that ``at least half of my neighbors are active,'' while in the second network no node will make this observation. Thus, even though only three of the 14 nodes are active, it appears to all inactive nodes in the first network that most of their neighbors are active.

The ``majority illusion'' can dramatically impact social contagions. Researchers use the threshold model to describe the spread of social contagions in networks~\cite{Granovetter78,Watts02,Centola07b}. At each time step in this model, an inactive individual observes the current states of its $k$ neighbors, and becomes active if more than $\phi k$ of the neighbors are active; otherwise, it remains inactive. The fraction $0 \le \phi \le 1$ is the activation \emph{threshold}. It represents the amount of social proof an individual requires before switching to the active state~\cite{Granovetter78}. Threshold of $\phi = 0.5$ means that to become active, an individual has to have a majority neighbors in the active state.
Though the two networks in Figure~\ref{fig:net14} have the same topology, when the threshold is $\phi=0.5$, all nodes will eventually become active in the network on the left, but not in the network on the right. This is because the ``majority illusion'' alters local neighborhoods of the nodes, distorting their perceptions of the prevalence of the active state.

This paper describes and analyzes the ``majority illusion'' paradox. We measure the strength of the paradox as the fraction of network nodes with a majority active neighbors. Using synthetic and real-world networks, we study how network structure and configuration of active nodes contributes to the paradox. We demonstrate empirically, as well as through theoretical analysis, that the paradox is stronger in networks in which the better-connected nodes are active, and also in networks with a heterogeneous degree distribution. Network structure also amplifies the paradox via degree correlations. The paradox is strongest in networks where low degree nodes have the tendency to connect to high degree nodes. Activating the high degree nodes in such networks biases the local observations of many nodes, which in turn impacts collective phenomena emerging in networks, including social contagions. Our statistical model quantifies the strength of this effect.


\section{Results}
\label{sec:results}
A network's structure is partly specified by its degree distribution $p(k)$, which gives the probability that a randomly chosen node in an undirected network has $k$ neighbors (i.e., degree $k$). This quantity also affects the probability that a randomly chosen edge is connected to a node of degree $k$, otherwise known as neighbor degree distribution $q(k)$. Since high degree nodes have more edges, they will be over-represented in the neighbor degree distribution by a factor proportional to their degree; hence, $q(k)= kp(k)/\avg k$, where $\avg k$ is the average node degree.

Networks often have structure beyond that specified by  their degree distribution: for example, nodes may preferentially link to others with a similar (or very different) degree.
Such degree correlation is captured by the joint degree distribution $e(k,\kpr)$, the probability to find nodes of degrees $k$ and $\kpr$ at either end of a randomly chosen edge in an undirected network~\cite{newman2002}. This quantity obeys normalization conditions $\sum_{k \kpr}{e(k,\kpr)}=1$ and $\sum_{\kpr} e(k,\kpr)=q(k)$. Globally, degree correlation in an undirected network is quantified by the assortativity coefficient, which is simply the Pearson correlation between degrees of connected nodes:
\begin{eqnarray*}
\label{eq:assortativity}
r&=&\frac{1}{\sigma_{q}^2}\sum_{k,\kpr}k\kpr \left[e(k,\kpr)-q(k)q(\kpr)\right] \\
&=& \frac{1}{\sigma_{q}^2}\left[\left(\sum_{k,\kpr} k\kpr e(k,\kpr)\right) - \avg k_q^2 \right].
\end{eqnarray*}
Here, $\sigma^2_q= \sum_k{k^2 q(k)} - \left[ \sum_k{k q(k)}\right]^2$. In assortative networks ($r > 0$), nodes have a tendency link to similar nodes, e.g., high-degree nodes to other high-degree nodes. In disassortative networks ($r < 0$), on the other hand, they prefer to link to dissimilar nodes. A star composed of a central hub and nodes linked only to the hub is an example of a disassortative network.

We can use Newman's edge rewiring procedure~\cite{newman2002} to change a network's assortativity without changing its degree distribution $p(k)$. The rewiring procedure randomly chooses two pairs of connected nodes and swaps their edges if doing so changes their degree correlation. This can be repeated until desired assortativity is achieved.

The configuration of attributes in a network is specified by the joint probability distribution $P(x,k)$, the probability that node of degree $k$ has an attribute $x$. In this work, we consider binary attributes only, and refer to nodes with $x=1$ as active and those with $x=0$ as inactive.
The joint distribution can be used to compute $\rho_{kx}$, the correlation between node degrees and attributes:
\begin{eqnarray}
\label{eq:rho}
\rho_{kx} & \equiv & \frac{1}{\sigma_x\sigma_k}\sum_{x,k}xk\left[P(x,k)-P(x)p(k)\right] \\ \nonumber
& = & \frac{1}{\sigma_x\sigma_k}\sum_{k}k\left[P(x=1,k)-P(x=1)p(k)\right] \\ \nonumber
& = &  \frac{P(x=1)}{\sigma_x\sigma_k}\left[\avg{k}_{x=1}-\avg{k}\right].
\end{eqnarray}
In the equations above, $\sigma_k$ and $\sigma_x$ are the standard deviations of the degree and attribute distributions respectively, and $\avg{k}_{x=1}$ is the average degree of active nodes.

Randomly activating nodes creates a configuration with $\rho_{kx}$ close to zero. We can change it by swapping attribute values among the nodes. For example, to increase $\rho_{kx}$, we randomly choose nodes $v_1$ with $x=1$ and $v_0$ with $x=0$ and swap their attributes if the degree of $v_0$ is greater than the degree of $v_1$.
We can continue swapping attributes until desired $\rho_{kx}$ is achieved (or it no longer changes).

\subsection{``Majority Illusion'' in Synthetic and Real-world Networks}
\label{sec:empirical}

\begin{figure*}[htbp] 
\centering
\begin{tabular}{ccc}
\subfigure[$\alpha=2.1$\label{fig:alpha2.1}]
{\includegraphics[width=\figwidth]{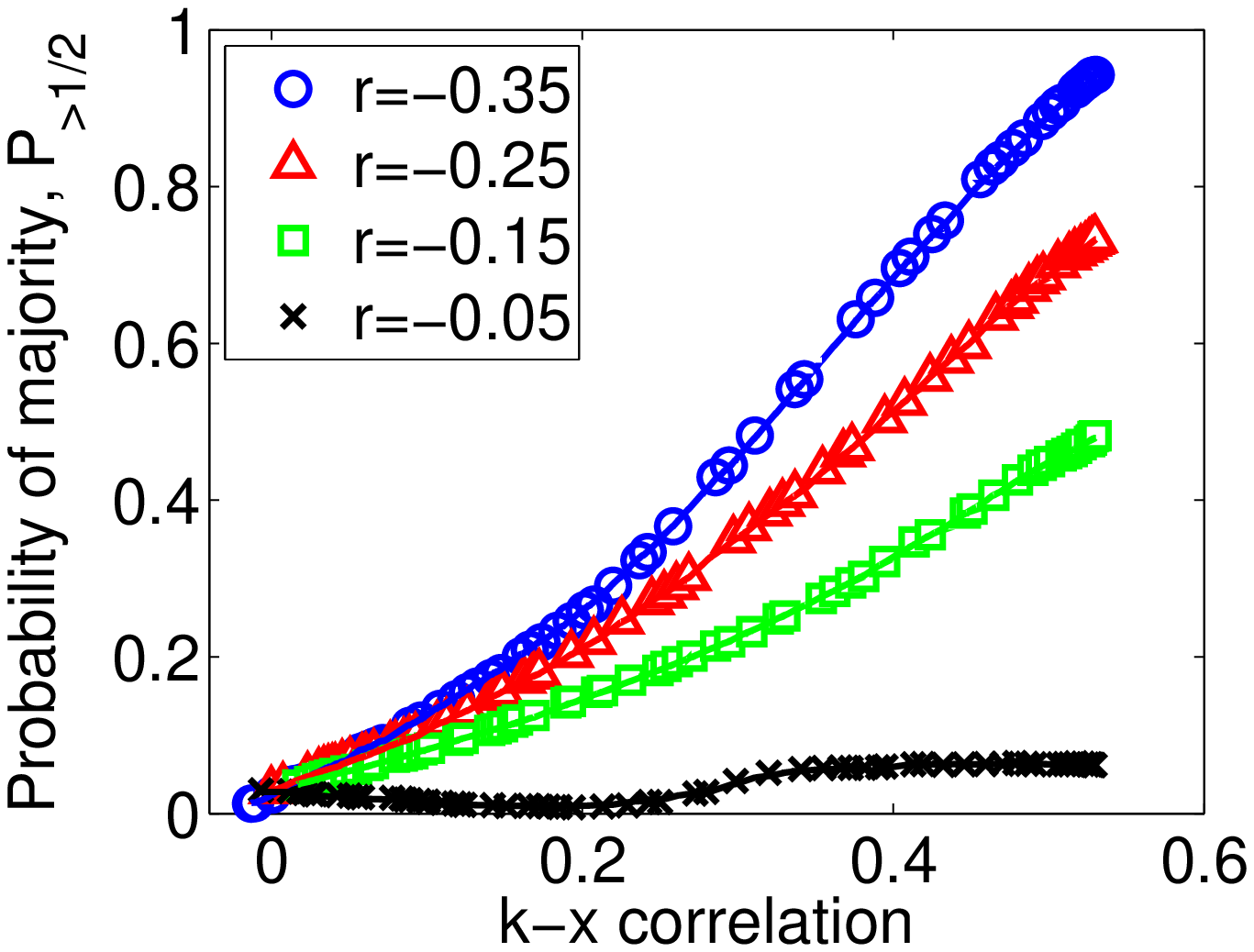}}
&
\subfigure[$\alpha=2.4$\label{fig:alpha2.4}]
{\includegraphics[width=\figwidth]{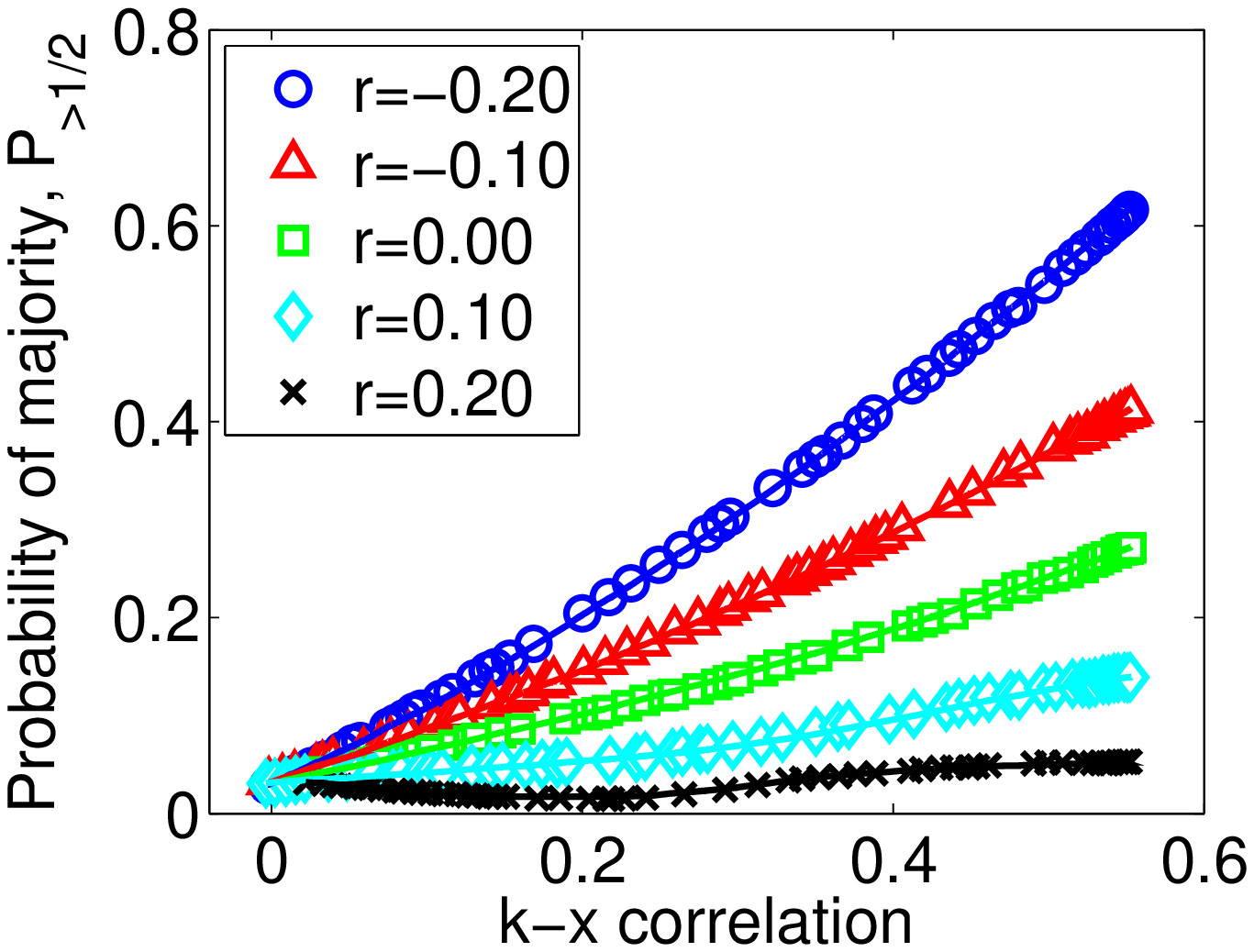}}
&
\subfigure[$\alpha=3.1$\label{fig:alpha3.1}]
{\includegraphics[width=\figwidth]{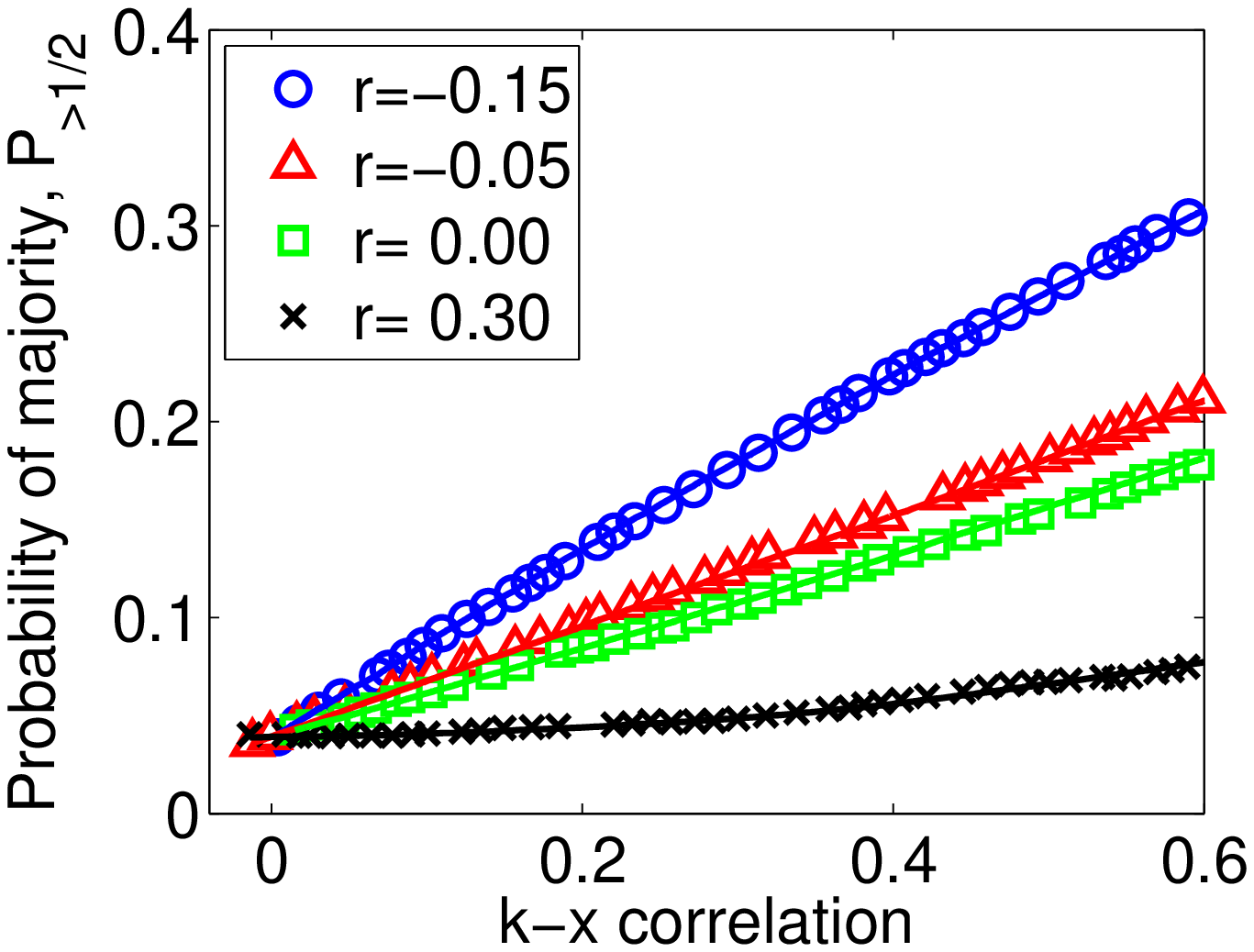}}
\end{tabular}
\caption{Magnitude of the ``majority illusion'' in scale-free networks as a function of degree--attribute correlation $\rho_{kx}$ and for different values of degree assortativity $r$. Each network has 10,000 nodes and degree distribution of the form $p(k) \sim k^{-\alpha}$. The fraction of active nodes in all cases is 5\%. The lines represent calculations using the statistical model of Equation~\protect\ref{eq:p>phi}. }
    \label{fig:scalefree}
 \end{figure*}

Synthetic networks allow us to systematically study how network structure affects the strength of the ``majority illusion'' paradox. First, we looked at scale-free networks. We generated networks with $N=10,000$ nodes and degree distribution of the form $p(k) \sim k^{-\alpha}$. Such networks are used to model the heterogeneous structure of many real-world networks, which contain a few high degree hubs and many low degree nodes. To create a scale-free network, we first sampled a degree sequence from a distribution with exponent $\alpha$, where exponent $\alpha$ took three different values (2.1, 2.4, and 3.1), and then used the configuration model to create an undirected network with that degree sequence (see \sectA{sec:data}).
We activated $P(x=1)=0.05$ of nodes and used edge rewiring and attribute swapping procedures describe above to change the network's degree assortativity $r$ and degree--attribute correlation $\rho_{kx}$.

Figure~\ref{fig:scalefree} shows the fraction of nodes with a majority active neighbors in these scale-free networks as a function of the degree--attribute correlation $\rho_{kx}$ and for different values of degree assortativity $r$. The fraction of nodes experiencing the ``majority illusion'' can be quite large. For $\alpha=2.1$, 60\%--80\% of the nodes may observe a majority active neighbors, even though only 5\% of the nodes are, in fact, active.
The ``majority illusion'' is exacerbated by three factors: it becomes stronger as the degree--attribute correlation increases, and as the network becomes more disassortative (i.e., $r$ decreases) and heavier-tailed (i.e., $\alpha$ becomes smaller). However, even when $\alpha=3.1$, under some conditions a substantial fraction of nodes will experience the paradox.
The lines in the figure show show theoretical estimates of the paradox using Equation~\ref{eq:p>phi}, as described in the next subsection.

\begin{figure*}[htbp] 
\centering
\begin{tabular}{ccc}
\multicolumn{3}{c}{Erd\H{o}s-R\'{e}nyi network with $N=10,000$ and $\avg k=5.2$}\\
\subfigure[$P(x=1)=0.05$\label{fig:ERa-5}]{\includegraphics[width=\figwidth]{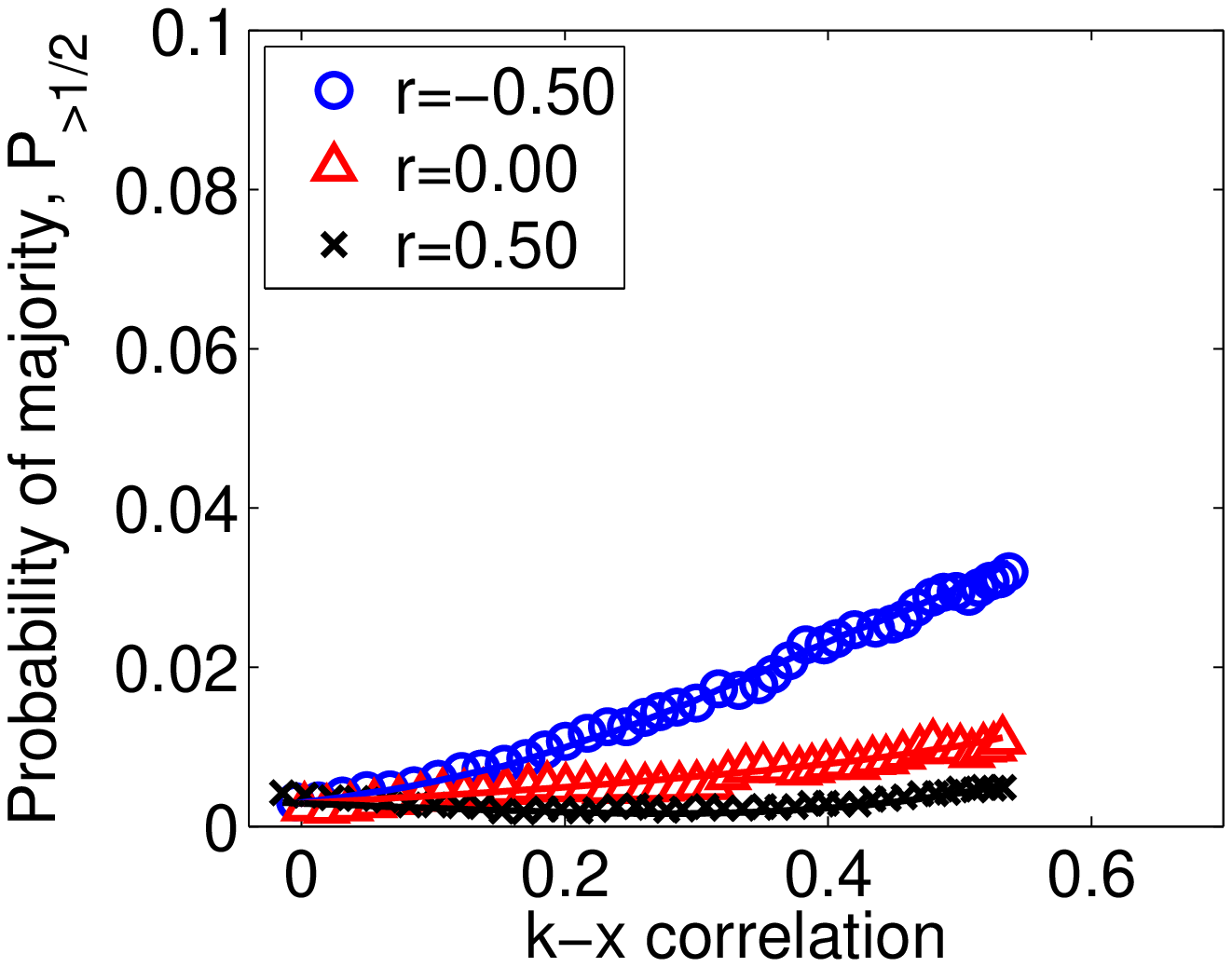}}
&
\subfigure[$P(x=1)=0.10$\label{fig:ERa-10}]{\includegraphics[width=\figwidth]{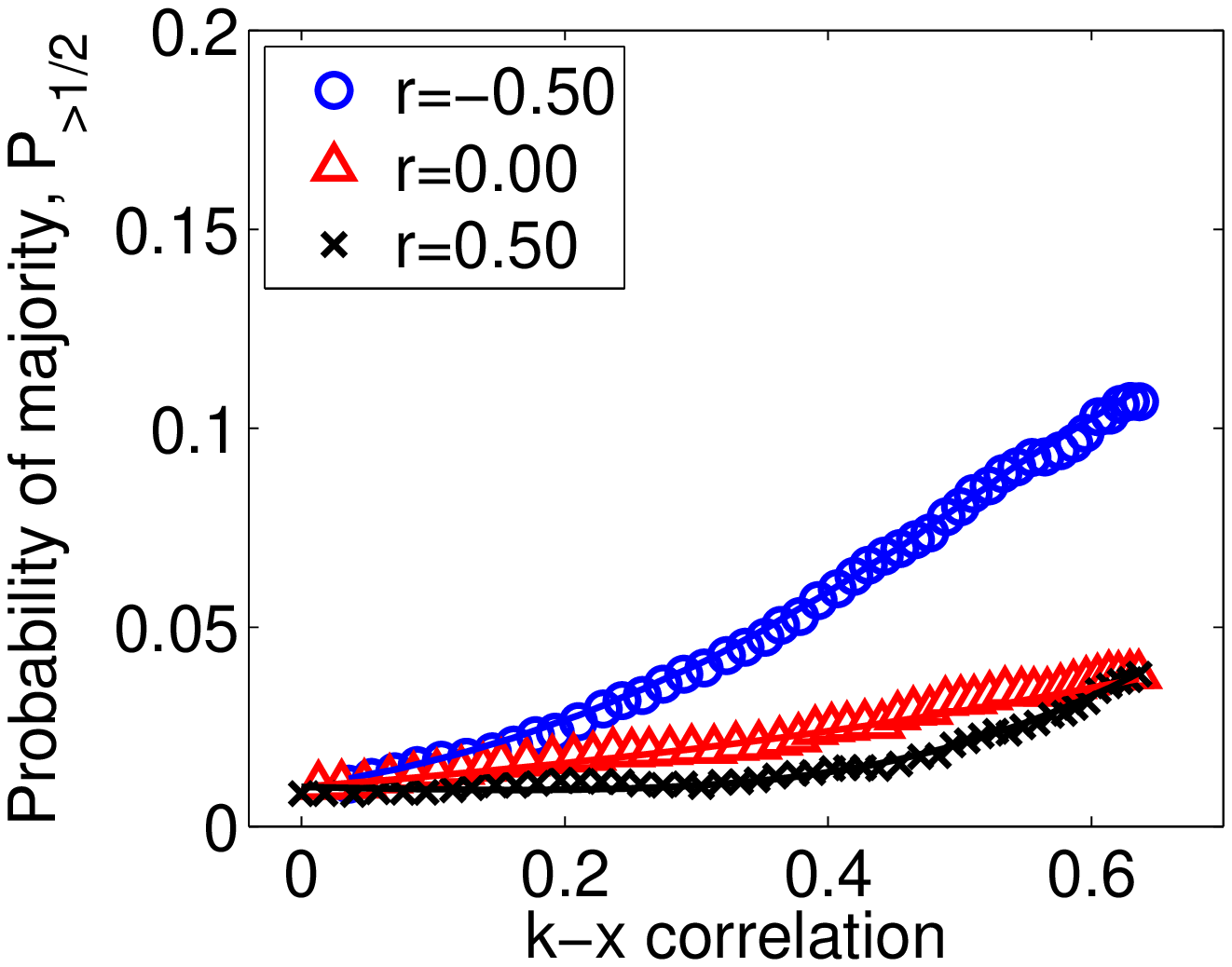}}
&
\subfigure[$P(x=1)=0.20$\label{fig:ERa-20}]{\includegraphics[width=\figwidth]{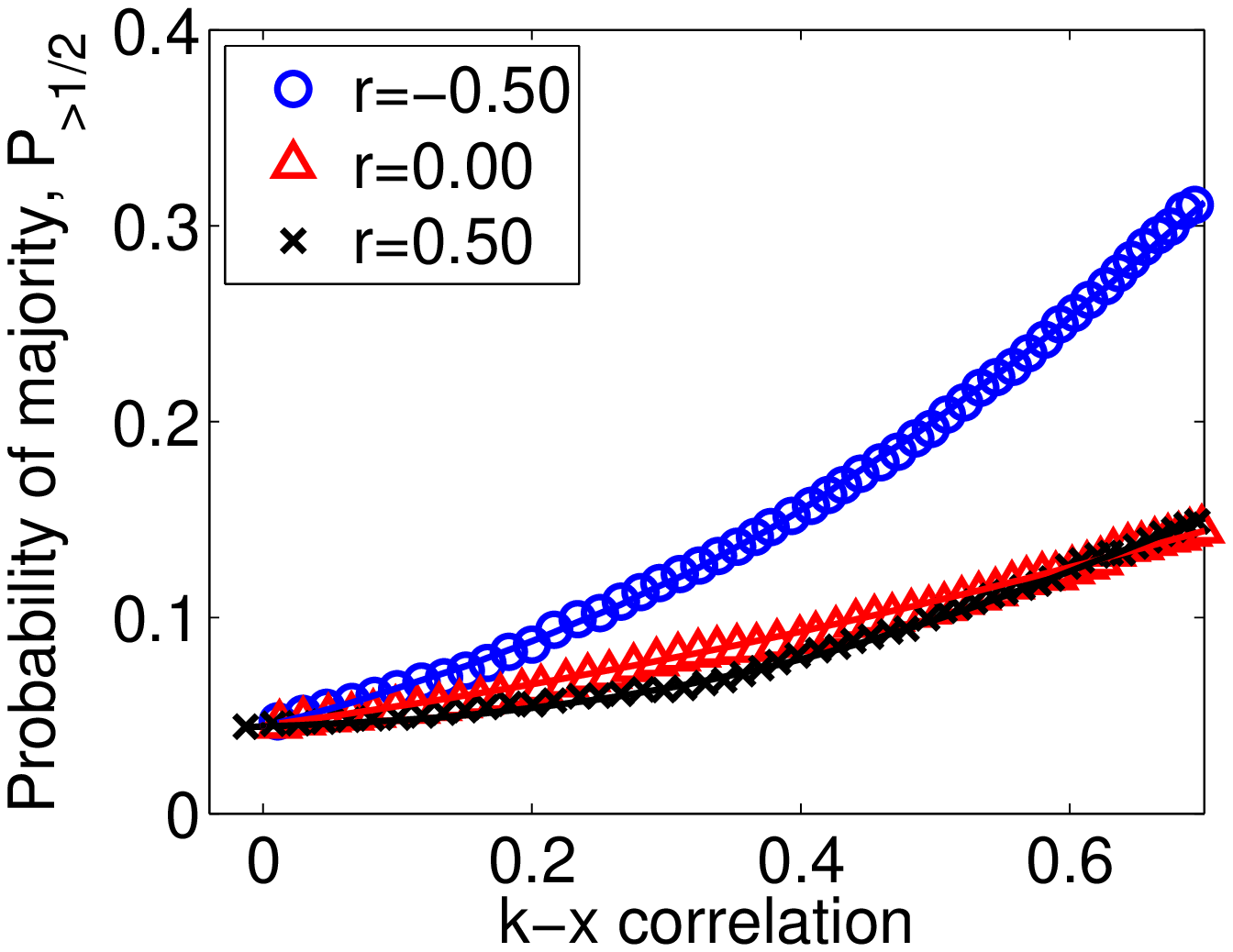}}
\\
\multicolumn{3}{c}{Erd\H{o}s-R\'{e}nyi network with $N=10,000$  and $\avg k=2.5$}\\
\subfigure[$P(x=1)=0.05$\label{fig:ERb-5}]{\includegraphics[width=\figwidth]{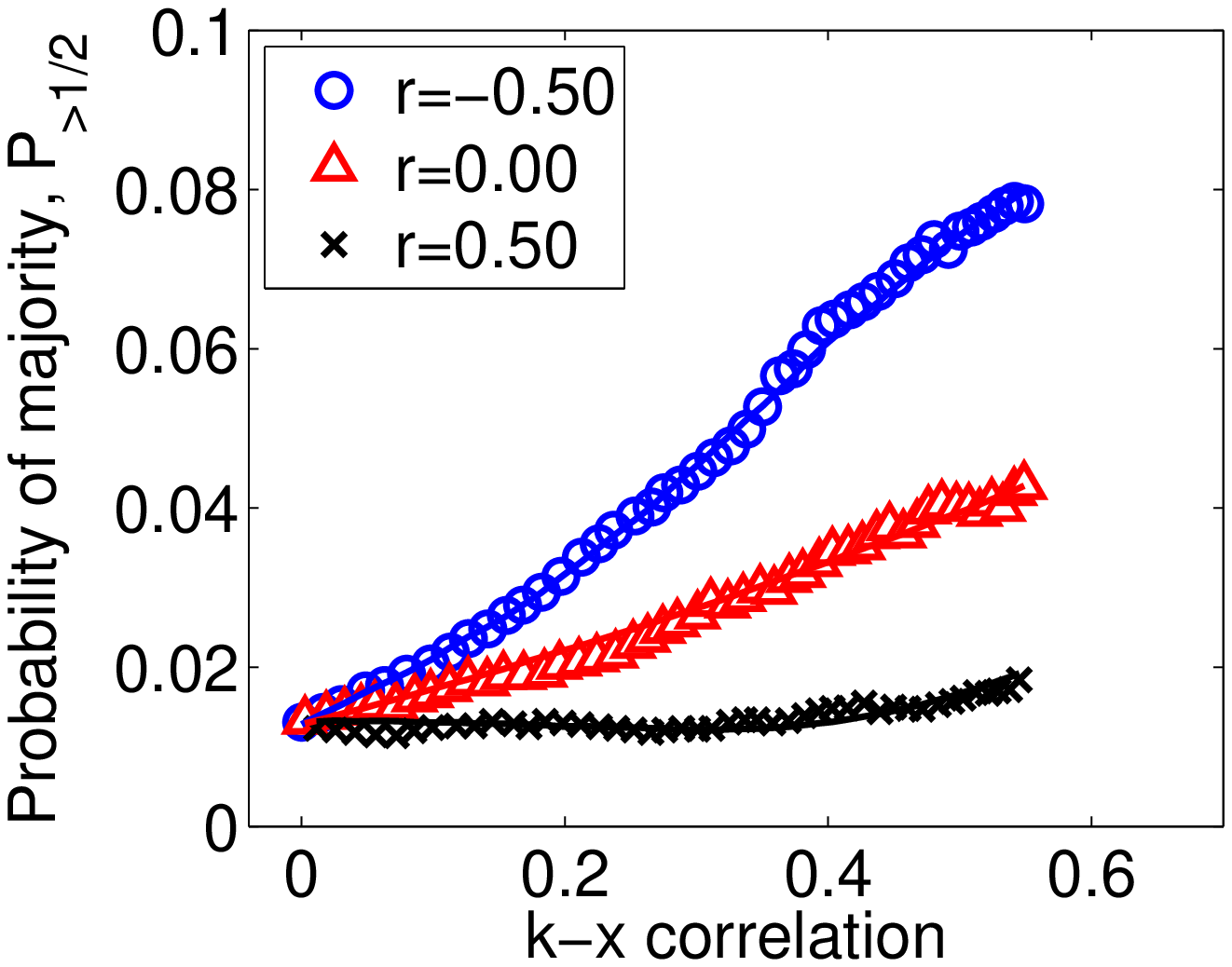}}
 &
\subfigure[$P(x=1)=0.10$\label{fig:ERb-10}]{\includegraphics[width=\figwidth]{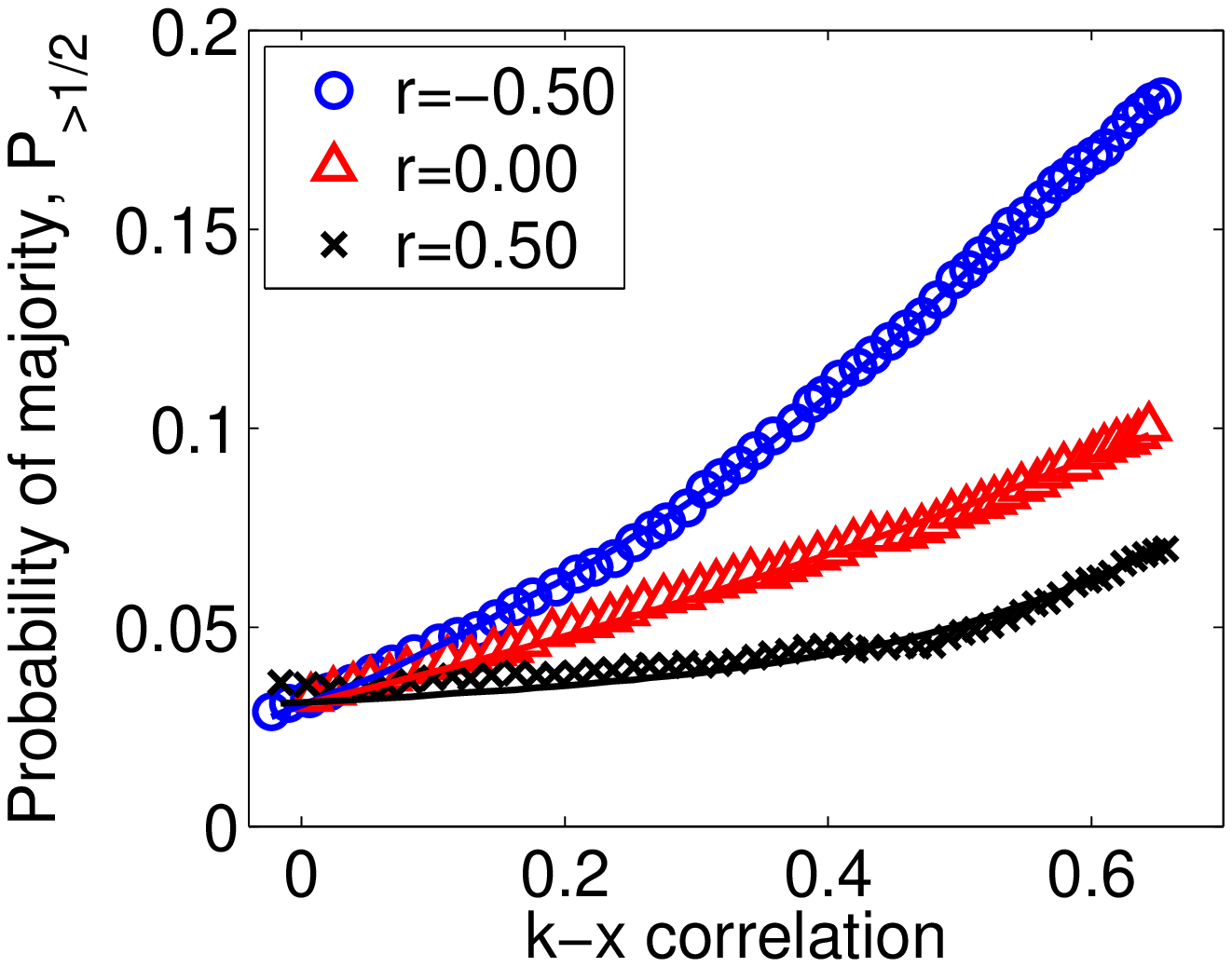}}
 &
\subfigure[$P(x=1)=0.20$\label{fig:ERb-20}]{\includegraphics[width=\figwidth]{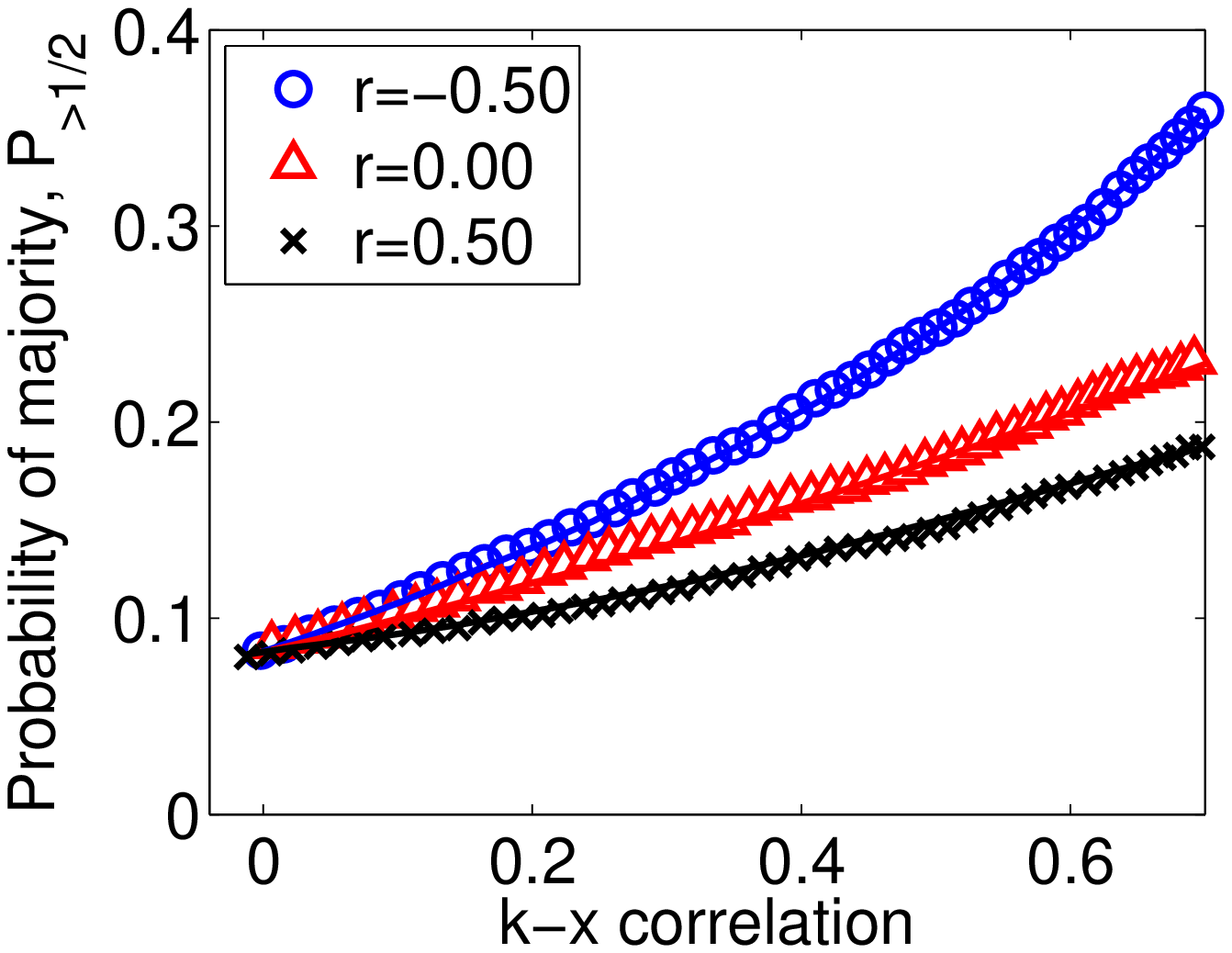}}
\end{tabular}
\caption{Magnitude of the ``majority illusion'' in Erd\H{o}s-R\'{e}nyi-type networks as a function of degree--attribute correlation $\rho_{kx}$ and for different values of degree assortativity $r$. Each network has 10,000 nodes with $\avg k=5.2$ (top row) or $\avg k=2.5$ (bottom row), and different fractions of active nodes. The lines represent calculations using the statistical model of Equation~\protect\ref{eq:p>phi}. }
    \label{fig:random}
\end{figure*}

``Majority illusion'' can also be observed in networks with a Poisson degree distribution. We used the Erd\H{o}s-R\'{e}nyi model to generate networks with $N=10,000$ and average degrees $\avg k = 5.2$ and $\avg k = 2.5$ (see \sectA{sec:data}). We randomly activated 5\%, 10\%, and 20\% of the nodes, and used edge rewiring and attribute swapping to change $r$ and $\rho_{kx}$ in these networks. Figure~\ref{fig:random} shows the fraction of nodes in the paradox regime. Though much reduced compared to scale-free networks, we still observe some amount of the paradox, especially in networks with a greater fraction of active nodes.

\begin{figure*}[htbp] 
\centering
\begin{tabular}{ccc}
{\includegraphics[width=\figwidth]{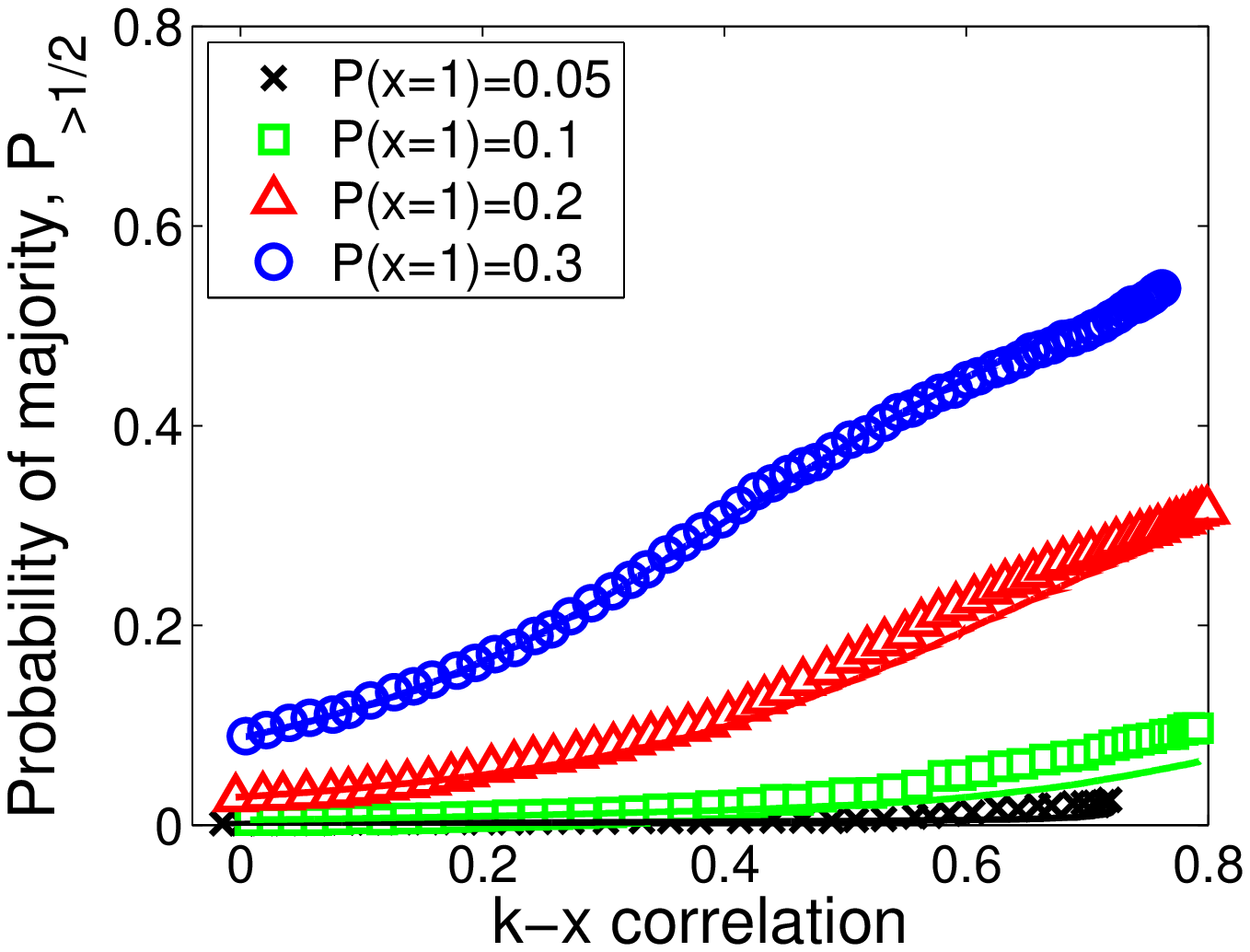}}
 &
{\includegraphics[width=\figwidth]{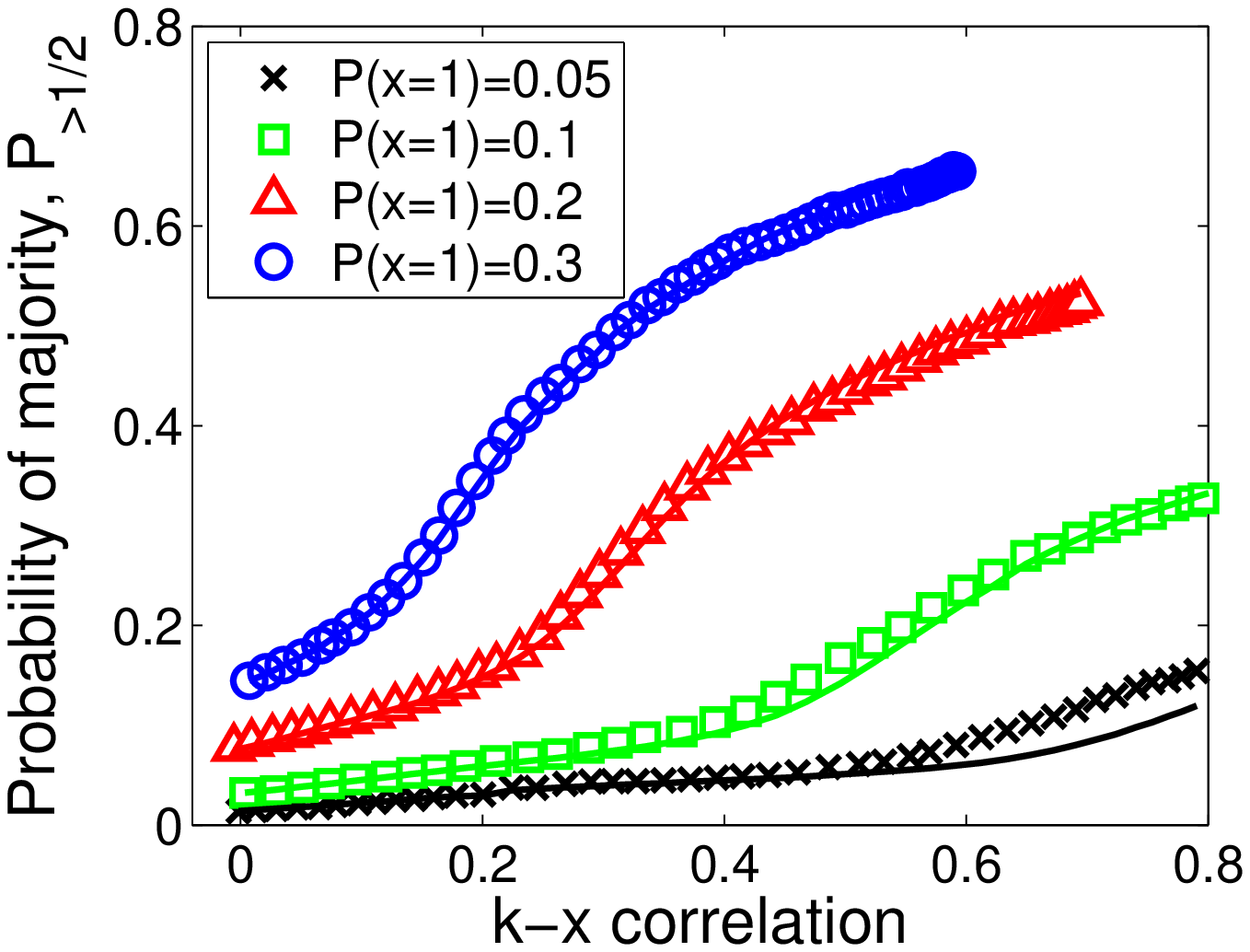}}
 &
{\includegraphics[width=\figwidth]{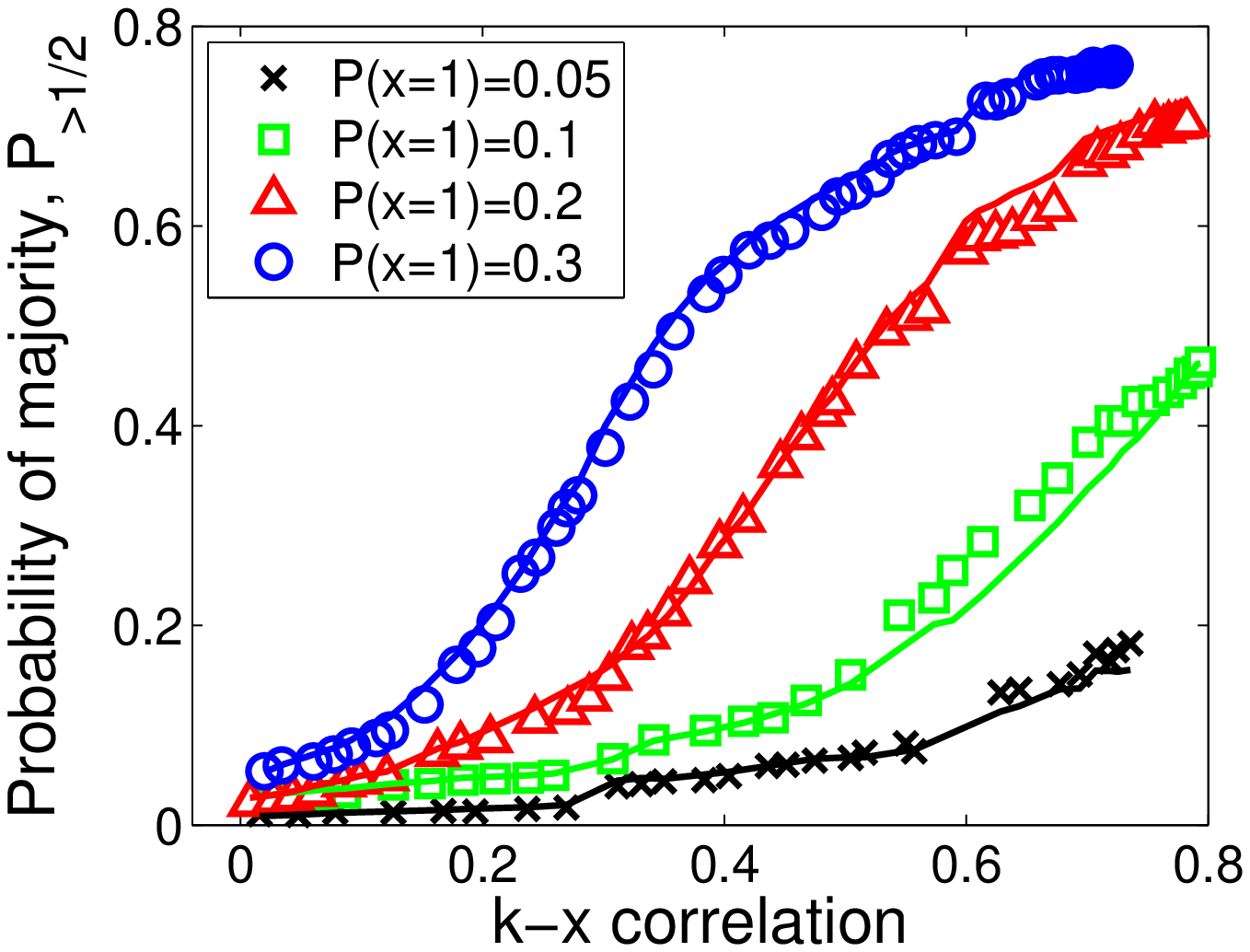}}
\\
HepTh collaboration & Digg  & blogs
\end{tabular}
\caption{Magnitude of the ``majority illusion'' in real-world networks as a function of degree--attribute correlation $\rho_{kx}$ for different fraction of active nodes $P(x=1)$. The lines represent calculations using the statistical model of Equation~\protect\ref{eq:p>phi}.}
    \label{fig:real-world}
 \end{figure*}

We also examined whether ``majority illusion'' can be manifested in real-world networks. We looked at three typical social and communications networks: the co-authorship network of high energy physicists (HepTh)~\cite{leskovec2007graph}, social media follower graph (Digg)~\cite{Hogg12epj}, and the network representing links between political blogs (blogs)~\cite{adamic2005political}.
All three networks are undirected. To make the Digg follower graph undirected, we kept only the mutual follow links, and further reduced the graph by extracting the largest connected component. Except for political blogs ($r=-0.22$), the networks were assortative with $r=0.23$ and $r=0.12$ for HepTh  and Digg respectively. These data are described in \sectA{sec:data}.

Figure~\ref{fig:real-world} shows the fraction of nodes experiencing the ``majority illusion'' for different fractions of active nodes $P(x=1)=0.05,\ 0.1,\ 0.2$ and $0.3$. As degree--attribute correlation $\rho_{kx}$ increases, a substantial fraction of nodes experience the paradox in almost all networks. The effect is largest in the disassortative political blogs network, where for high enough correlation,  as many as 60\%--70\% of nodes will have a majority active neighbors, even when only 20\% of the nodes are active. The effect also exists in the Digg network of mutual followers, and to a lesser degree in the HepTh co-authorship network.
Although positive assortativity reduces the magnitude of the effect, compared with synthetic networks, local perceptions of nodes in real-world networks can also be skewed. If the attribute represents an opinion, under some conditions, even a minority opinion can appear to be extremely popular locally.

\subsection{Modeling ``Majority Illusion'' in Networks}
\label{sec:quantify}
Having demonstrated empirically some of the relationships between ``majority illusion'' and network structure, we next develop a model that includes network properties in the calculation of paradox strength.
Like the friendship paradox, the ``majority illusion'' is rooted in differences between degrees of nodes and their neighbors~\cite{Feld91,Gupta15sbp}.
These differences result in nodes perceiving that, not only are their neighbors better connected~\cite{Feld91}, on average, but that they also have more of some attribute than they  themselves have~\cite{Hodas13icwsm}. The latter paradox, which is referred to as the generalized friendship paradox, is enhanced by correlations between node degrees and attribute values $\rho_{kx}$~\cite{Eom14}. In binary attribute networks, where nodes can be either active or inactive, a configuration in which higher degree nodes tend to be active causes the remaining nodes to observe that their neighbors are more active than they are (see \sectA{sec:friendship}).

While heterogeneous degree distribution and degree--attribute correlations give rise to friendship paradoxes even in random networks, other elements of network structure, such as correlation between degrees of connected nodes, may also affect observations nodes make of their neighbors. To understand why, consider the probability that a node has an active neighbor, conditioned on that node having a degree $k$:
\begin{eqnarray}
\label{eq:p(x'|k)}
 P(\xpr=1|k) &=& \sum_{\kpr} P(\xpr=1|\kpr)P(\kpr|k) \\ \nonumber
 &=& \sum_{\kpr} P(\xpr=1|\kpr)\frac{e(k,\kpr)}{q(k)}.
\end{eqnarray}
In the equation above, $e(k,\kpr)$ is the joint degree distribution. Globally, the probability that any node has an active neighbor is
\begin{align*}
P(\xpr=1) &= \sum_k P(\xpr=1|k)p(k) \\
&= \sum_k\sum_{\kpr}P(\xpr=1|\kpr)\frac{e(k,\kpr)}{q(k)}p(k)\\
&= \sum_k \sum_{\kpr}\frac{P(\xpr=1, \kpr)}{p(\kpr)}e(k,\kpr)\frac{\avg k}{k} \\
&= \sum_{\kpr}\frac{P(\xpr=1,\kpr)}{q(\kpr)} \sum_k \frac{\kpr}{k}e(k,\kpr).
\end{align*}
Given two different networks with the same degree distribution $p(k)$ and the same configuration of active nodes, the probability that a node in each network observes an active neighbor $P(\xpr=1)$ is a function of $\sum_{k,\kpr}(\kpr/{k}) {e(k,\kpr)}$. Since assortativity $r$ is a function of $\sum_{k,k'}kk'e(k,k')$,  we can see that the two expressions weigh the  $e(k,\kpr)$ term in opposite ways.
This suggests that the probability of having an active neighbor increases as assortativity decreases and vice versa. Thus, we expect stronger paradoxes in disassortative networks.

To quantify the ``majority illusion'' paradox, we calculate the probability that a node of degree $k$ has more than a fraction $\phi$ of active neighbors, i.e., neighbors with attribute value $\xpr=1$:
\begin{eqnarray}
\label{eq:p>phi(k)}
P_{>\phi} (k) &=& \sum_{n>\phi k}^k {k\choose n} \times \\ \nonumber
 && P(\xpr=1|k)^n (1-P(\xpr=1|k))^{(k-n)}\;.
\end{eqnarray}
Here $P(\xpr=1|k)$ is the conditional probability of having an active neighbor, given a node with degree $k$, and is specified by Eq.~\ref{eq:p(x'|k)}. Although the threshold $\phi$ in Eq.~\ref{eq:p>phi(k)} could be any fraction, in this paper we focus on $\phi=\frac{1}{2}$, which represents a straight majority. Thus, the fraction of all nodes most of whose neighbors are active is
\begin{eqnarray}
\label{eq:p>phi}
P_{>\frac{1}{2}} &=& \sum_k p(k) \sum_{n>\frac{k}{2}}^k {k\choose n} \times \\ \nonumber
 && P(\xpr=1|k)^n (1-P(\xpr=1|k))^{(k-n)}\;.
\end{eqnarray}

Using Equation~\ref{eq:p>phi}, we can calculate the strength of the ``majority illusion'' paradox for any network whose degree sequence, joint degree distribution $e(k,\kpr)$, and conditional attribute distribution $P(x|k)$ are known. The solid lines in Figures~\ref{fig:scalefree}--\ref{fig:real-world} report these calculations for each network.  We used the empirically determined joint probability distribution $P(x,k)$ to calculate $P(x=1|k)$ in the equation above, as well as $\rho_{kx}$ for the figure.
For some ``well-behaved'' degree distributions, $P(x=1|k)$ can be determined analytically (rather than empirically) by approximating the joint distribution $P(x,k)$ as a bivariate normal distribution (see \sectA{sec:gaussian}). However, this ``gaussian'' approximation breaks down as degree distributions becomes heavier tailed.
Overall, our statistical model  does a good job explaining most of the empirical observations.
Although the global degree assortativity $r$ is an important contributor to the ``majority illusion,'' a more detailed view of the structure using joint degree distribution $e(k,\kpr)$ is necessary to accurately estimate the magnitude of the paradox. As demonstrated in \sectA{sec:structure}, two networks with the same $p(k)$ and $r$ can display different amounts of the paradox.

\section{Discussion}
Local prevalence of some attribute among a node's network neighbors can be very different from its global prevalence, creating an illusion that the attribute is far more common than it actually is. In a social network, this illusion may cause people  to reach wrong conclusions about how common a behavior is, leading them to accept as a norm a behavior that is globally rare. This may explain how global outbreaks can be triggered by very few initial adopters, and why people overestimate how much their friends engage in risky behaviors, such as alcohol and drug use.

We quantified this paradox, which we call the ``majority illusion'', and studied its dependence on network structure and attribute configuration. As in the friendship paradox~\cite{Feld91,Hodas13icwsm,Eom14,Kooti14icwsm}, ``majority illusion'' can ultimately be traced to the power of high degree nodes to skew the observations of many others. This is because such nodes are overrepresented in the local neighborhoods of other nodes. This, by itself is not surprising, given than high degree nodes are expected to have more influence and are often targeted by influence maximization algorithms~\cite{Kempe03}. However, the ability of high degree nodes to bias the perceptions of others depends on other aspects of network structure. Specifically, we showed that the paradox is much stronger in disassortative networks, where high degree nodes tend to link to low degree nodes. In other words, given the same degree distribution, the high degree nodes in a disassortative network will have greater power to skew the observations of others than those in an assortative network. This suggests that some network structures are more susceptible than others to influence manipulation and the spread of
external shocks~\cite{Watts02}.
Furthermore, small changes in network topology, assortativity and degree--attribute correlation may further exacerbate the paradox even when there are no actual changes in the distribution of the attribute. This may explain the apparently sudden shifts in public attitudes witnessed during the Arab Spring and on the question of gay marriage.

The ``majority illusion'' is an example of class size bias effect. When sampling data to estimate average class or event size, more popular classes and events will be overrepresented in the sample, biasing estimates of their average size~\cite{Feld77}. Thus, the average class size that students experience at college is larger than the college's average class size. Similarly, people experience highways, restaurants, and events to be more crowded than they normally are. In networks, sampling bias affects estimates of network structure, including its degree distribution~\cite{Achlioptas06,Gupta15sbp}. Our work suggests that network bias also affects an individual's local perceptions and the collective  social phenomena that emerge.


\subsection*{Acknowledgements}
Authors are grateful to Nathan Hodas and Farshad Kooti for their inputs into this work.


\begin{thebibliography}{10}

\bibitem{Achlioptas06}
Dimitris Achlioptas, Aaron Clauset, David Kempe, and Cristopher Moore.
\newblock {On the Bias of Traceroute Sampling; or, Power-law Degree
  Distributions in Regular Graphs}.
\newblock In {\em {Proc. 37th ACM Symposium on Theory of Computing (STOC)}},
  2005.

\bibitem{adamic2005political}
Lada~A Adamic and Natalie Glance.
\newblock The political blogosphere and the 2004 us election: divided they
  blog.
\newblock In {\em Proceedings of the 3rd international workshop on Link
  discovery}, pages 36--43. ACM, 2005.

\bibitem{Baer91}
J.~S. Baer, A.~Stacy, and M.~Larimer.
\newblock {Biases in the perception of drinking norms among college students.}
\newblock {\em Journal of studies on alcohol}, 52(6):580--586, November 1991.

\bibitem{Bearak14}
Jonathan~M. Bearak.
\newblock {Casual Contraception in Casual Sex: {Life-Cycle} Change in
  Undergraduates' Sexual Behavior in Hookups}.
\newblock {\em Social Forces}, 93(2):sou091--513, October 2014.

\bibitem{bender1978asymptotic}
Edward~A Bender and E~Rodney Canfield.
\newblock {The asymptotic number of labeled graphs with given degree
  sequences}.
\newblock {\em Journal of Combinatorial Theory, Series A}, 24(3):296--307,
  1978.

\bibitem{berkowitz2005overview}
Alan~D Berkowitz.
\newblock {An overview of the social norms approach}.
\newblock {\em Changing the culture of college drinking: A socially situated
  health communication campaign}, pages 193--214, 2005.

\bibitem{Bettencourt06}
Lu{\'i}s Bettencourt, Ariel Cintron-Arias, David~I Kaiser, and Carlos
  Castillo-Chavez.
\newblock {The power of a good idea: Quantitative modeling of the spread of
  ideas from epidemiological models}.
\newblock {\em Physica A: Statistical Mechanics and its Applications},
  364:513--536, 2006.

\bibitem{Centola10}
Damon Centola.
\newblock {The Spread of Behavior in an Online Social Network Experiment}.
\newblock {\em Science}, 329(5996):1194--1197, September 2010.

\bibitem{Centola15}
Damon Centola and Andrea Baronchelli.
\newblock The spontaneous emergence of conventions: An experimental study of
  cultural evolution.
\newblock {\em Proceedings of the National Academy of Sciences},
  112(7):1989--1994, February 2015.

\bibitem{Centola07b}
Damon Centola, V{\'i}ctor~M. Egu{\'i}luz, and Michael~W. Macy.
\newblock {Cascade dynamics of complex propagation}.
\newblock {\em Physica A: Statistical Mechanics and its Applications},
  374(1):449--456, January 2007.

\bibitem{Christakis10}
Nicholas~A. Christakis and James~H. Fowler.
\newblock {Social Network Sensors for Early Detection of Contagious Outbreaks}.
\newblock {\em PLoS ONE}, 5(9):e12948+, September 2010.

\bibitem{Cohen03}
Reuven Cohen, Shlomo Havlin, and Daniel ben Avraham.
\newblock {Efficient Immunization Strategies for Computer Networks and
  Populations}.
\newblock {\em Phys. Rev. Lett.}, 91:247901, Dec 2003.

\bibitem{Eom14}
Young-Ho Eom and Hang-Hyun Jo.
\newblock {Generalized friendship paradox in complex networks: The case of
  scientific collaboration}.
\newblock {\em Scientific Reports}, 4, April 2014.

\bibitem{Feld91}
Scott~L. Feld.
\newblock {Why Your Friends Have More Friends Than You Do}.
\newblock {\em American Journal of Sociology}, 96(6):1464--1477, May 1991.

\bibitem{Feld77}
Scott~L. Feld and Bernard Grofman.
\newblock {Variation in Class Size, the Class Size Paradox, and Some
  Consequences for Students}.
\newblock {\em Research in Higher Education}, 6(3), 1977.

\bibitem{GarciaHerranz14}
Manuel Garcia-Herranz, Esteban Moro, Manuel Cebrian, Nicholas~A. Christakis,
  and James~H. Fowler.
\newblock {Using Friends as Sensors to Detect Global-Scale Contagious
  Outbreaks}.
\newblock {\em PLoS ONE}, 9(4):e92413+, April 2014.

\bibitem{Granovetter78}
Mark Granovetter.
\newblock {Threshold Models of Collective Behavior}.
\newblock {\em American Journal of Sociology}, 83(6):1420--1443, 1978.

\bibitem{Gupta15sbp}
Sidharth Gupta, Xiaoran Yan, and Kristina Lerman.
\newblock {Structural Properties of Ego Networks}.
\newblock In {\em {International Conference on Social Computing, Behavioral
  Modeling and Prediction}}, 2015.

\bibitem{Hodas13icwsm}
Nathan Hodas, Farshad Kooti, and Kristina Lerman.
\newblock {Friendship Paradox Redux: Your Friends Are More Interesting Than
  You}.
\newblock In {\em {Proc. 7th Int. AAAI Conf. on Weblogs And Social Media}},
  2013.

\bibitem{Hogg12epj}
Tad Hogg and Kristina Lerman.
\newblock Social dynamics of digg.
\newblock {\em EPJ Data Science}, 1(5), June 2012.

\bibitem{Kempe03}
David Kempe, Jon Kleinberg, and Eva Tardos.
\newblock {Maximizing the spread of influence through a social network}.
\newblock In {\em {KDD '03: Proceedings of the ninth ACM SIGKDD international
  conference on Knowledge discovery and data mining}}, pages 137--146, New
  York, NY, USA, 2003. ACM Press.

\bibitem{Kim15}
David~A. Kim, Alison~R. Hwong, Derek Stafford, D.~Alex Hughes, A.~James
  O'Malley, James~H. Fowler, and Nicholas~A. Christakis.
\newblock Social network targeting to maximise population behaviour change: a
  cluster randomised controlled trial.
\newblock {\em The Lancet}, May 2015.

\bibitem{Kooti14icwsm}
Farshad Kooti, Nathan~O. Hodas, and Kristina Lerman.
\newblock {Network Weirdness: Exploring the Origins of Network Paradoxes}.
\newblock In {\em {International Conference on Weblogs and Social Media
  (ICWSM)}}, March 2014.

\bibitem{leskovec2007graph}
Jure Leskovec, Jon Kleinberg, and Christos Faloutsos.
\newblock Graph evolution: Densification and shrinking diameters.
\newblock {\em ACM Transactions on Knowledge Discovery from Data (TKDD)},
  1(1):2, 2007.

\bibitem{Lloyd-Smith05}
J.~O. Lloyd-Smith, S.~J. Schreiber, P.~E. Kopp, and W.~M. Getz.
\newblock {Superspreading and the effect of individual variation on disease
  emergence}.
\newblock {\em Nature}, 438(7066):355--359, 2005.

\bibitem{molloy1995critical}
Michael Molloy and Bruce Reed.
\newblock {A critical point for random graphs with a given degree sequence}.
\newblock {\em Random structures \& algorithms}, 6(2-3):161--180, 1995.

\bibitem{newman2002}
M.~E.~J. Newman.
\newblock {Assortative Mixing in Networks}.
\newblock {\em Phys. Rev. Lett.}, 89:208701, Oct 2002.

\bibitem{Rogers03}
Everett~M. Rogers.
\newblock {\em {Diffusion of Innovations, 5th Edition}}.
\newblock Free Press, 5th edition, August 2003.

\bibitem{Salganik06}
M.~J. Salganik, P.~S. Dodds, and D.~J. Watts.
\newblock {Experimental Study of Inequality and Unpredictability in an
  Artificial Cultural Market}.
\newblock {\em Science}, 311, 2006.

\bibitem{Schelling73}
Thomas~C. Schelling.
\newblock {Hockey Helmets, Concealed Weapons, and Daylight Saving: A Study of
  Binary Choices with Externalities}.
\newblock {\em The Journal of Conflict Resolution}, 17(3), 1973.

\bibitem{Valente95book}
Thomas~W. Valente.
\newblock {\em {Network Models of the Diffusion of Innovations (Quantitative
  Methods in Communication Subseries)}}.
\newblock Hampton Press (NJ), 1995.

\bibitem{Watts02}
Duncan~J. Watts.
\newblock {A simple model of global cascades on random networks}.
\newblock {\em Proceedings of the National Academy of Sciences},
  99(9):5766--5771, April 2002.

\bibitem{Young11}
H.~Peyton Young.
\newblock The dynamics of social innovation.
\newblock {\em Proceedings of the National Academy of Sciences}, 108(Supplement
  4):21285--21291, December 2011.

\end{thebibliography}

\setcounter{figure}{0}
\makeatletter
\renewcommand{\thefigure}{S\@arabic\c@figure}
\makeatother

\setcounter{table}{0}
\makeatletter
\renewcommand{\thetable}{S\@arabic\c@table}
\makeatother

\section*{Appendix}
\setcounter{section}{0}
\makeatletter
\renewcommand{\thesection}{S\@arabic\c@section}
\makeatother

\section{Data}
\label{sec:data}
We use the configuration model~\cite{bender1978asymptotic,molloy1995critical}, as implemented by the SNAP library ({https://snap.stanford.edu/data/}), to create a scale-free network with a specified degree sequence. We generated a degree sequence from a power law of the form $p(k) \sim k^{-\alpha}$. Here, $p_k$ is fraction of nodes that have $k$ half-edges.  The configuration model proceeded by linking a pair of randomly chosen half-edges to form an edge. The linking procedure was repeated until all half-edges have been used up or there were no more ways to form an edge. 

To create Erd\H{o}s-R\'{e}nyi-type networks,  we started with $N=10,000$ nodes and linked pairs at random with some fixed probability. These probabilities were fixed to produce average degree similar to the average degree of the scale-free networks. 

The statistics of real-world networks we studied, including the collaboration network of high energy physicist (HepTh),$^1$ Digg follower graph ({http://www.isi.edu/$\sim$lerman/downloads/digg2009.html}),
and a network of political blogs ({http://www-personal.umich.edu/$\sim$mejn/netdata/}) are summarized below.

\small
\begin{tabular}{|r|c|c|c|c|}\hline
\emph{network} & \emph{nodes} & \emph{edges} & $\avg k$ & \emph{assortativity}\\\hline
HepTh & 9,877 & 25,998 & 5.3 & 0.2283 \\ \hline 
Digg & 25,454 & 175,892 & 13.8 & 0.1160 \\ \hline 
Political blogs & 1,490 & 19,090 & 25.6 & -0.2212\\ \hline 
\end{tabular}
\normalsize


\subsection{Friendship Paradox}
\label{sec:friendship}
Node degree distribution $p(k)$ gives the probability that a randomly chosen node in an undirected network has $k$ neighbors or edges. Neighbor degree distribution $q(k)$ gives the probability that a randomly chosen edge in an undirected network is connected to a node of degree $k$.
It is easy to demonstrate that average neighbor degree $\avg{k}_q$ is larger than average node degree $\avg{k}$. The difference between these quantities is
\begin{equation*}
\label{eq:friendship}
\avg{k}_q - \avg{k} = \sum_k \frac{k^2p(k)}{\avg k} - {\avg k} = \frac{\avg{k^2}- \avg{k}^2 }{\avg k} = \frac{\sigma^2_{k}}{\avg k} \,,
\end{equation*}
where $\sigma_{k}$ is the standard deviation of the degree distribution $p(k)$. Since $\sigma_k \ge 0$, $\avg{k}_q - \avg{k} \ge 0$. This confirms that the friendship paradox, which says that average neighbor degree is larger than node's own degree, has its origins in the heterogeneous degree distribution~\cite{Feld91}, and is more pronounced in networks with larger degree heterogeneity $\sigma_k$.

Heterogeneous degree distribution also contributes to nodes perceiving that 
their neighbors have more of some attribute than they  themselves have --- what is referred to as the generalized friendship paradox~\cite{Eom14}. Let's consider again a network where nodes have a binary attribute $x$. For convenience, we will refer to nodes with the attribute value $x=1$ as active, and those with $x=0$ as inactive. The probability that a random node is active is $P(x=1) = \sum_k P(x=1|k)p(k)$. The probability that a random neighbor is active is $Q(x=1) = \sum_k P(x=1|k)q(k) $. Using Bayes' rule, this can be rewritten as
\begin{eqnarray*}
Q(x=1) &=& \sum_k\frac{P(x=1,k)}{p(k)}\frac{kp(k)}{\avg k} \times \\ \nonumber
&=& \sum_k\frac{P(x=1,k)}{P(x=1)}\frac{kP(x=1)}{\avg k} \\ \nonumber
&=& \frac{P(x=1)}{\avg k}\sum_kkP(k|x=1) \\ \nonumber
&=& P(x=1)\frac{\avg k_{x=1}}{\avg k}\,,
\end{eqnarray*}
where $\avg k_{x=1}$ is the average degree of active nodes. This quantity and the average degree $\avg k$ are related via the correlation coefficient $\rho_{kx}=\frac{P(x=1)}{\sigma_x\sigma_k}\left[\avg{k}_{x=1}-\avg{k}\right]$ (Eq.~\ref{eq:rho}).
Hence, the strength of the generalized friendship paradox is
\begin{align*}
Q(x=1) - P(x=1) =  \rho_{kx}\frac{\sigma_x\sigma_k}{\avg k}\,,
\end{align*}
which is positive when node degree and attribute are positively correlated ($\rho_{kx} >0$) and increases with this correlation~\cite{Eom14}.

\subsection{Gaussian Approximation}
\label{sec:gaussian}
The conditional probability $P(x'=1|k')$ required to calculate the strength of the ``majority illusion'' using Eq.~\ref{eq:p>phi} can be specified analytically for some ``well-behaved'' degree distributions, such as scale--free distributions of the form $p(k)\sim k^{-\alpha}$ with $\alpha>3$ or the Poisson distributions of the Erd\H{o}s-R\'{e}nyi random graphs in near-zero assortativity. In these cases, probability $P(x'=1|k')$ can be acquired by approximating the joint distribution $P(x',k')$ as a multivariate normal distribution, and we have
\[
\avg{P(x'|k')}=\avg{P(x')} + \rho_{kx}\frac{\sigma_x}{\sigma_k}(k'-\avg k)\;, \]
resulting in
\[P(x'=1|k')=\avg x+\rho_{kx}\frac{\sigma_x}{\sigma_k}(k'-\avg k).
\]
\begin{figure*}[htbp] 
\centering
\begin{tabular}{ccc}
\subfigure[$\alpha=2.1$ \label{fig:gauss21}]{\includegraphics[width=\figwidth]{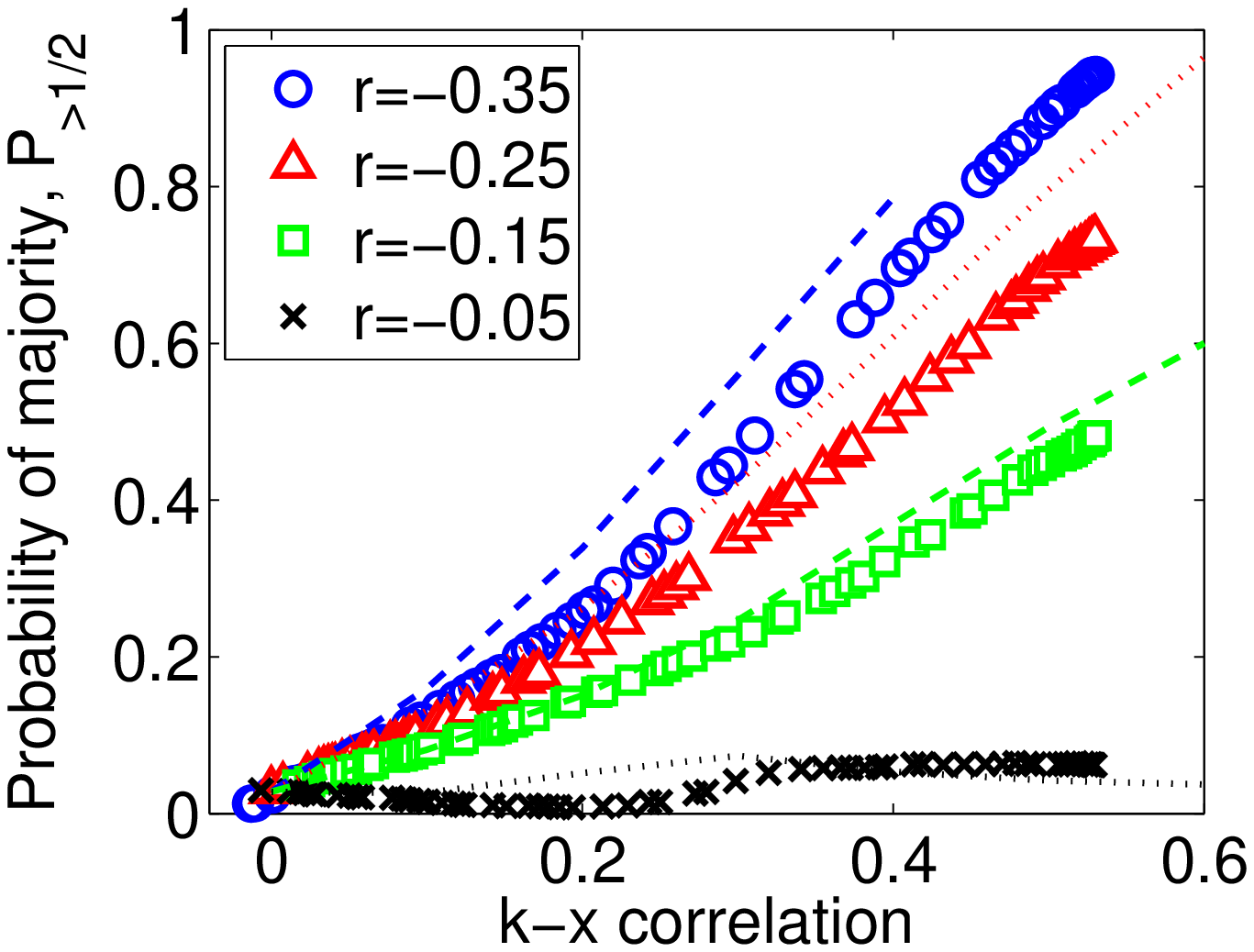}}
 &
\subfigure[$\alpha=2.4$ \label{fig:gauss24}]{\includegraphics[width=\figwidth]{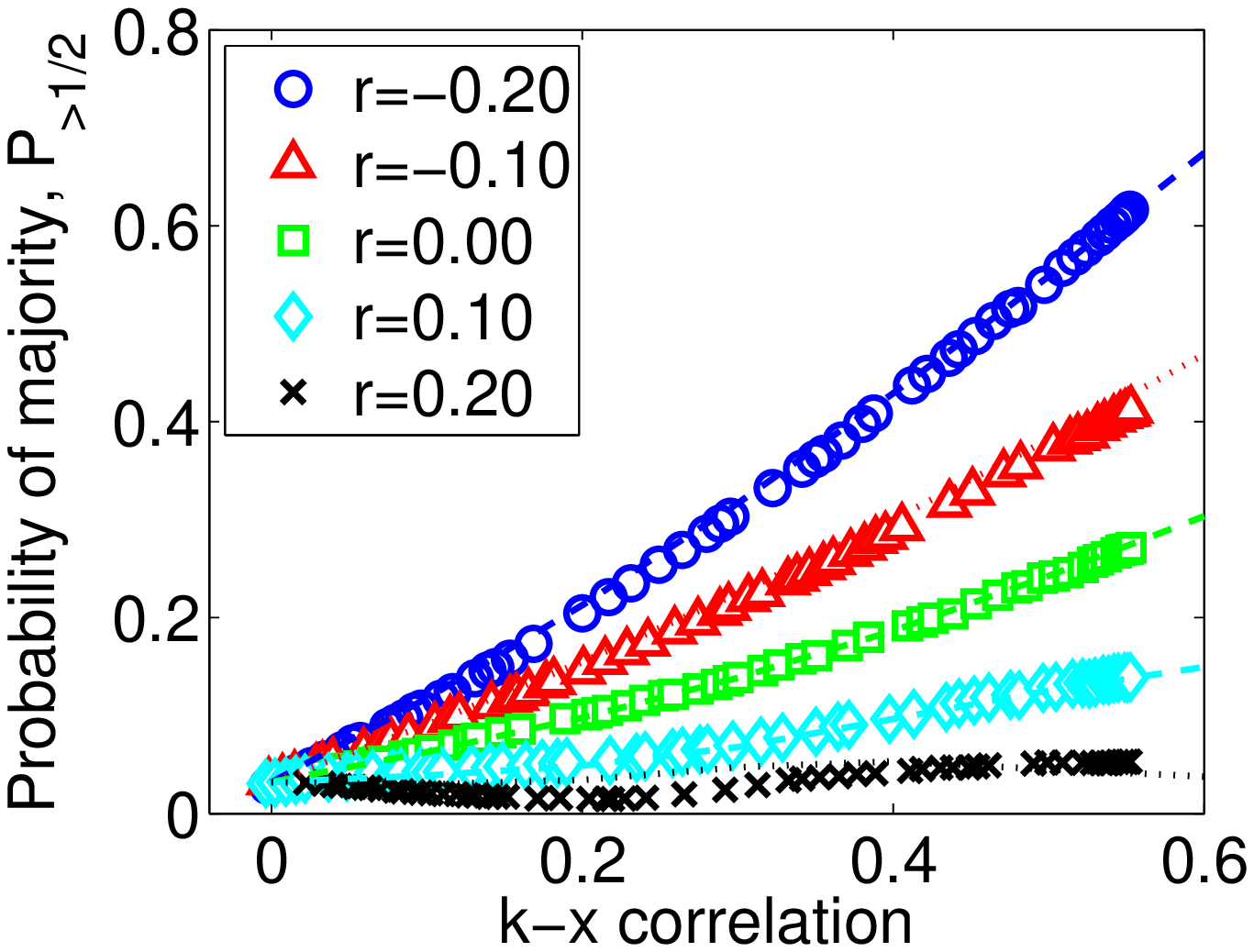}}
 &
\subfigure[$\alpha=3.1$ \label{fig:gauss31}]{\includegraphics[width=\figwidth]{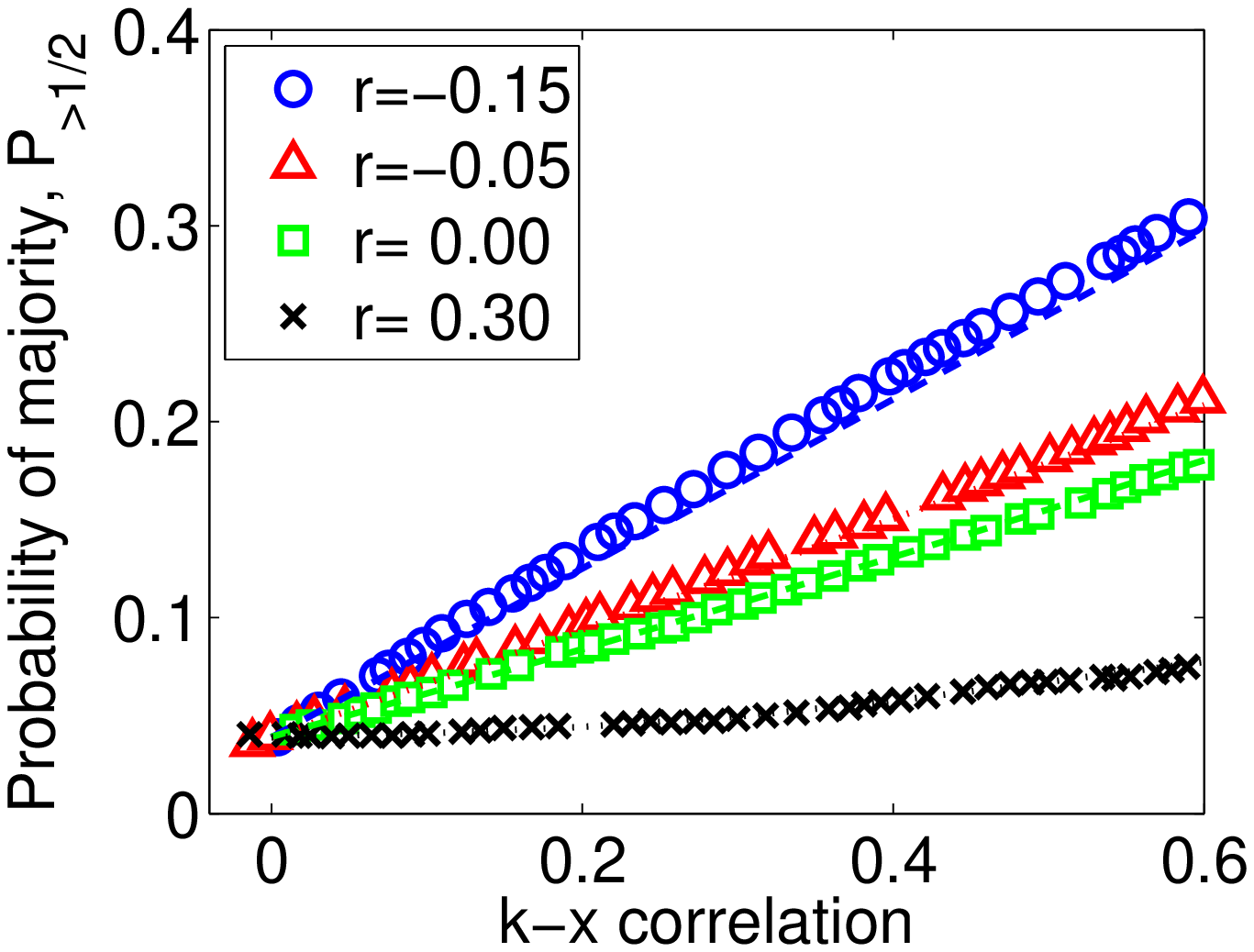}}
 \\
\subfigure[$\avg k=5.2$ \label{fig:gaussERa}]{\includegraphics[width=\figwidth]{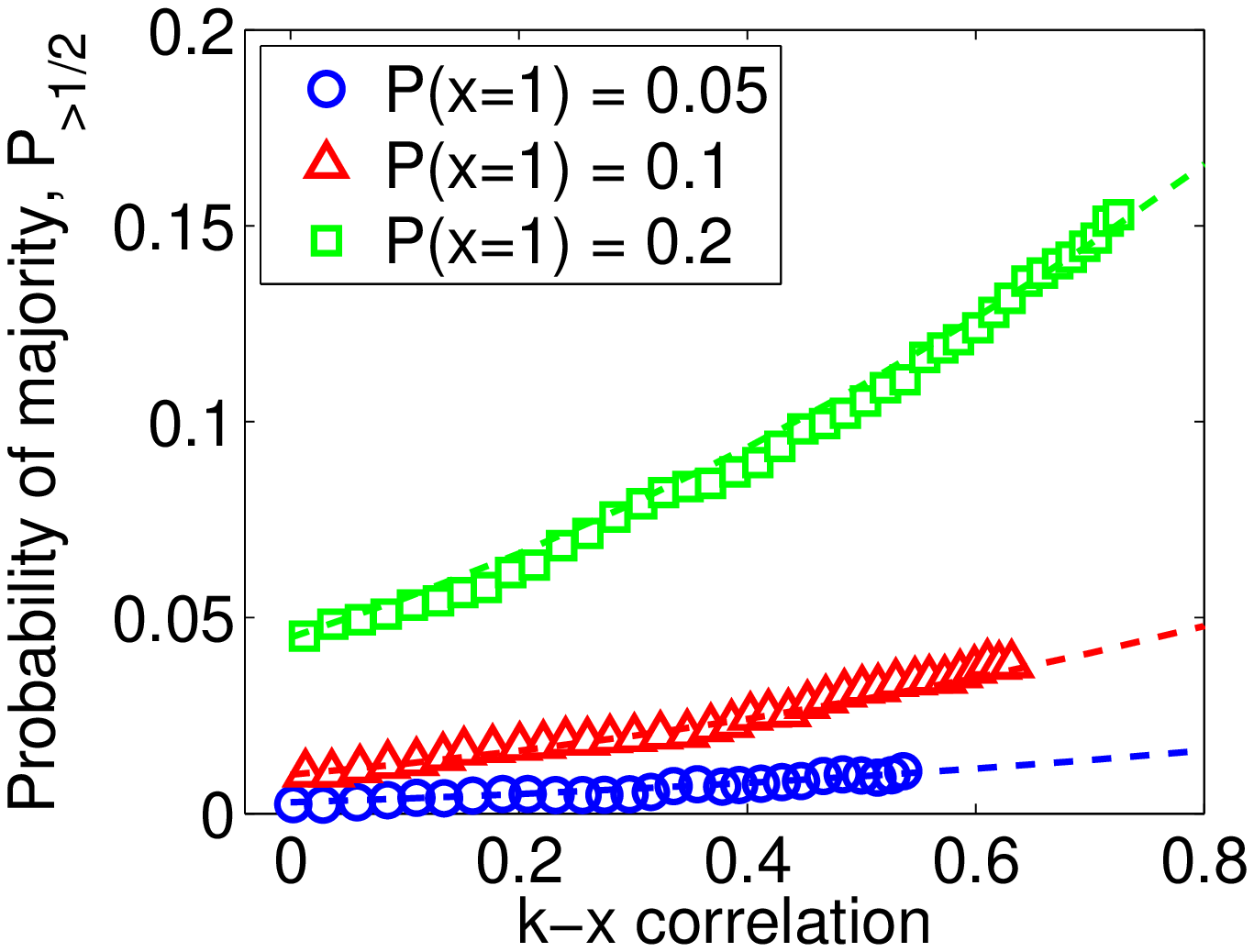}}
 &
\subfigure[$\avg k=2.5$ \label{fig:gaussERb}]{\includegraphics[width=\figwidth]{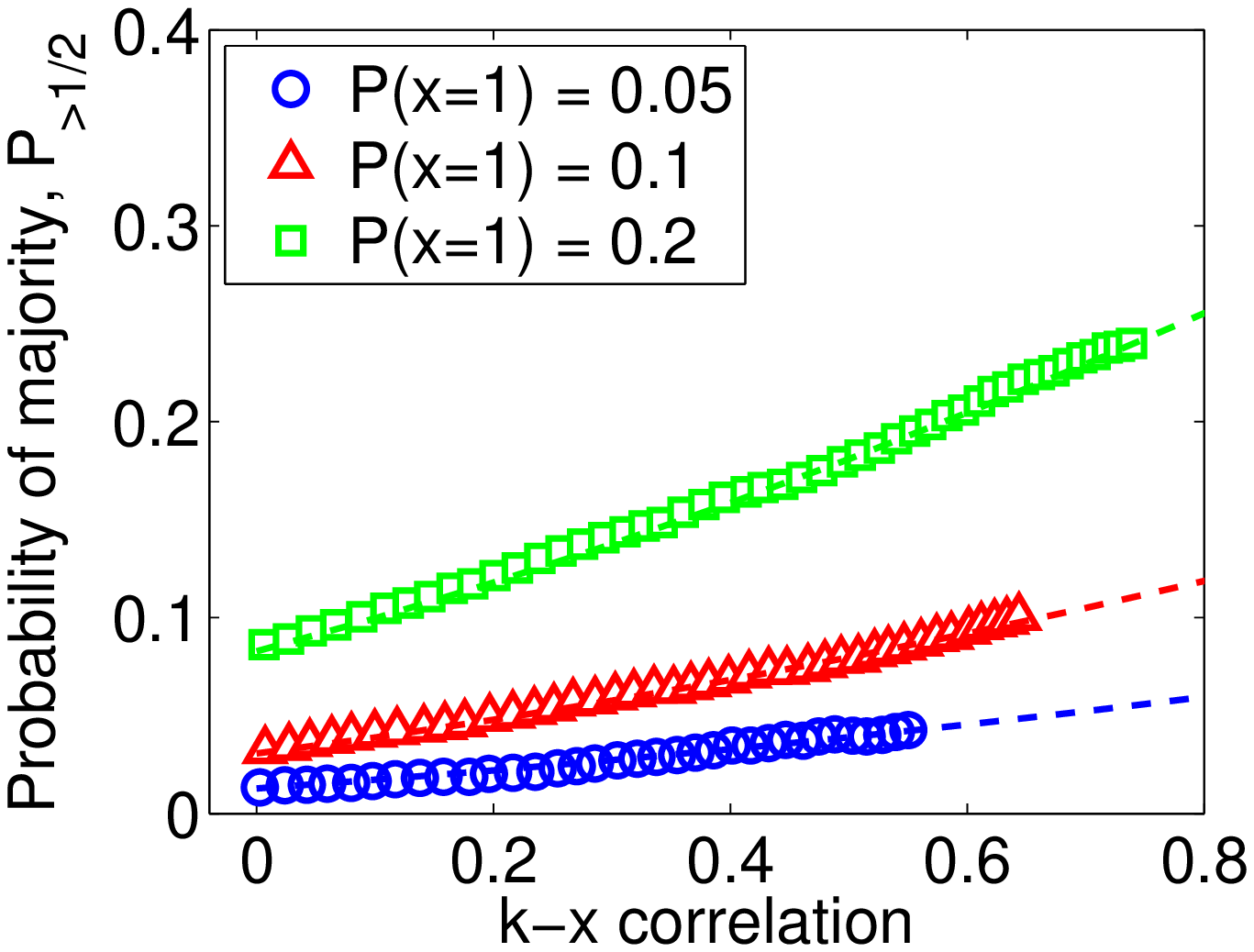}}
\end{tabular}
\caption{Strength of the ``majority illusion'' with Gaussian approximation. Symbols show the empirically determined fraction of nodes in the paradox regime (same as in Fig.~\protect\ref{fig:scalefree} and Fig.~\protect\ref{fig:random}), while dashed lines show theoretical estimates using the Gaussian approximation.}
    \label{fig:gauss}
\end{figure*}

Figure~\ref{fig:gauss} reports the ``majority illusion'' in the same synthetic scale--free networks as Fig.~\ref{fig:scalefree}, but with theoretical lines (dashed lines) calculated using the Gaussian approximation for estimating $P(x'=1|k')$. The Gaussian approximation fits results quite well for the network with degree distribution exponent $\alpha=3.1$. However, theoretical estimate deviates significantly from data in a network with a heavier--tailed degree distribution with exponent  $\alpha=2.1$. The approximation also deviate from the actual result when the network has very positive or negative assortativity. When this approximation is not applicable, we need to determine the joint distribution $P(x',k')$ empirically to calculate both $\rho_{kx}$ and $P(x'=1|k')$ for a specific joint distribution.

\subsection{Influence of Network Structure}
\label{sec:structure}
\begin{figure*}[htbp] 
\centering
\begin{tabular}{ccc}
\subfigure[$\alpha=2.1$ \label{fig:idseq21}]{\includegraphics[width=\figwidth]{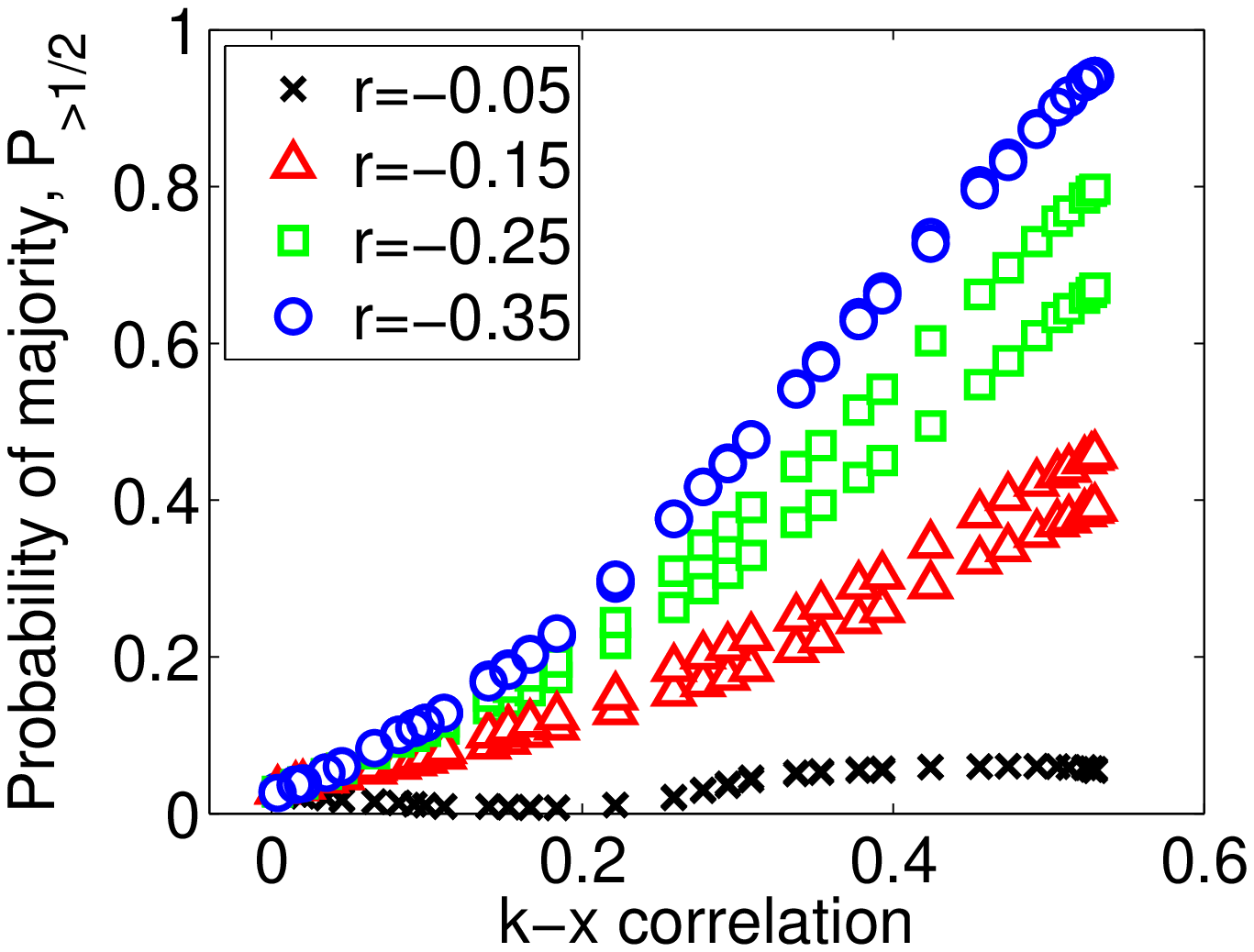}}
 &
\subfigure[$\alpha=2.4$ \label{fig:idseq24}]{\includegraphics[width=\figwidth]{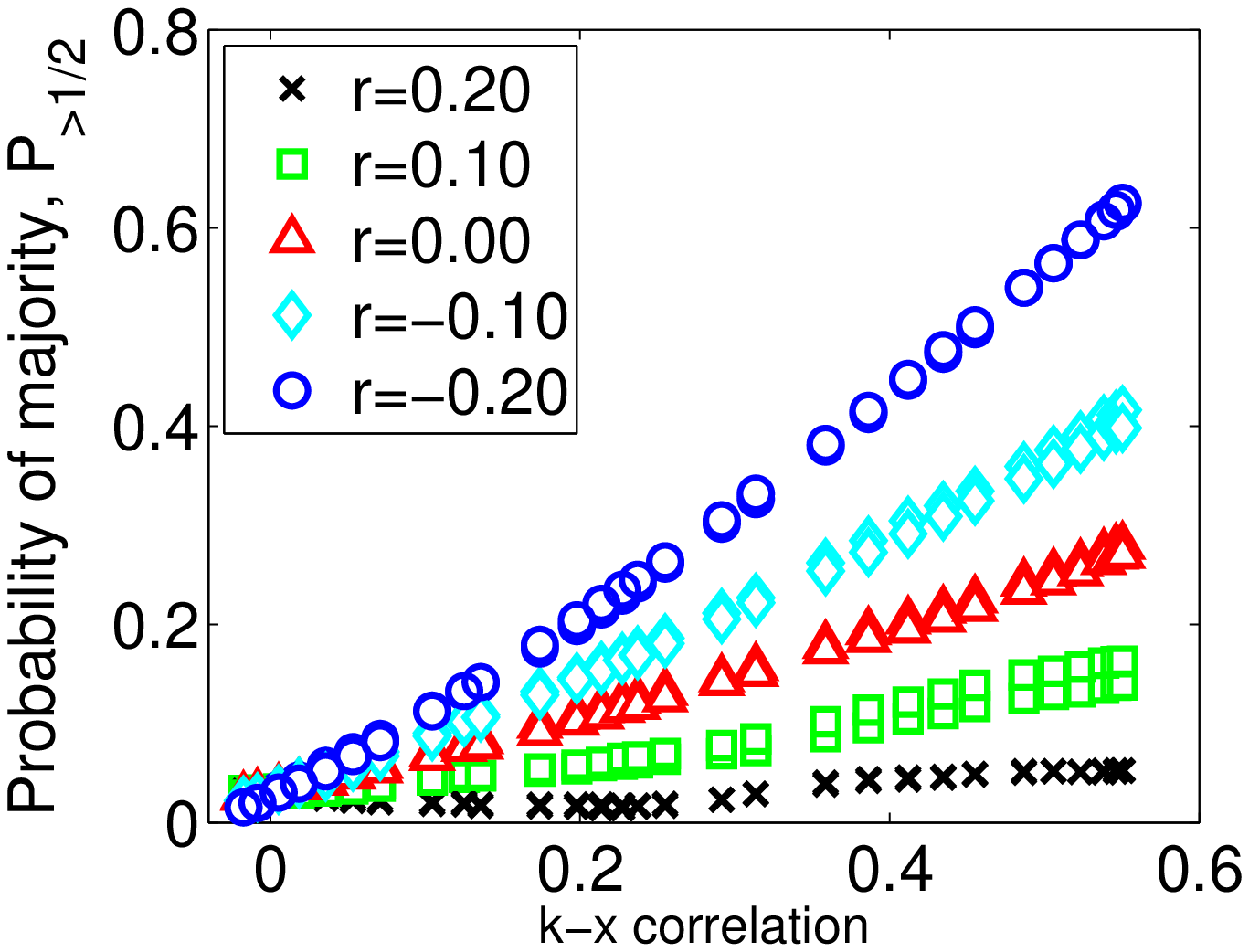}}
 &
\subfigure[$\alpha=3.1$ \label{fig:idseq31}]{\includegraphics[width=\figwidth]{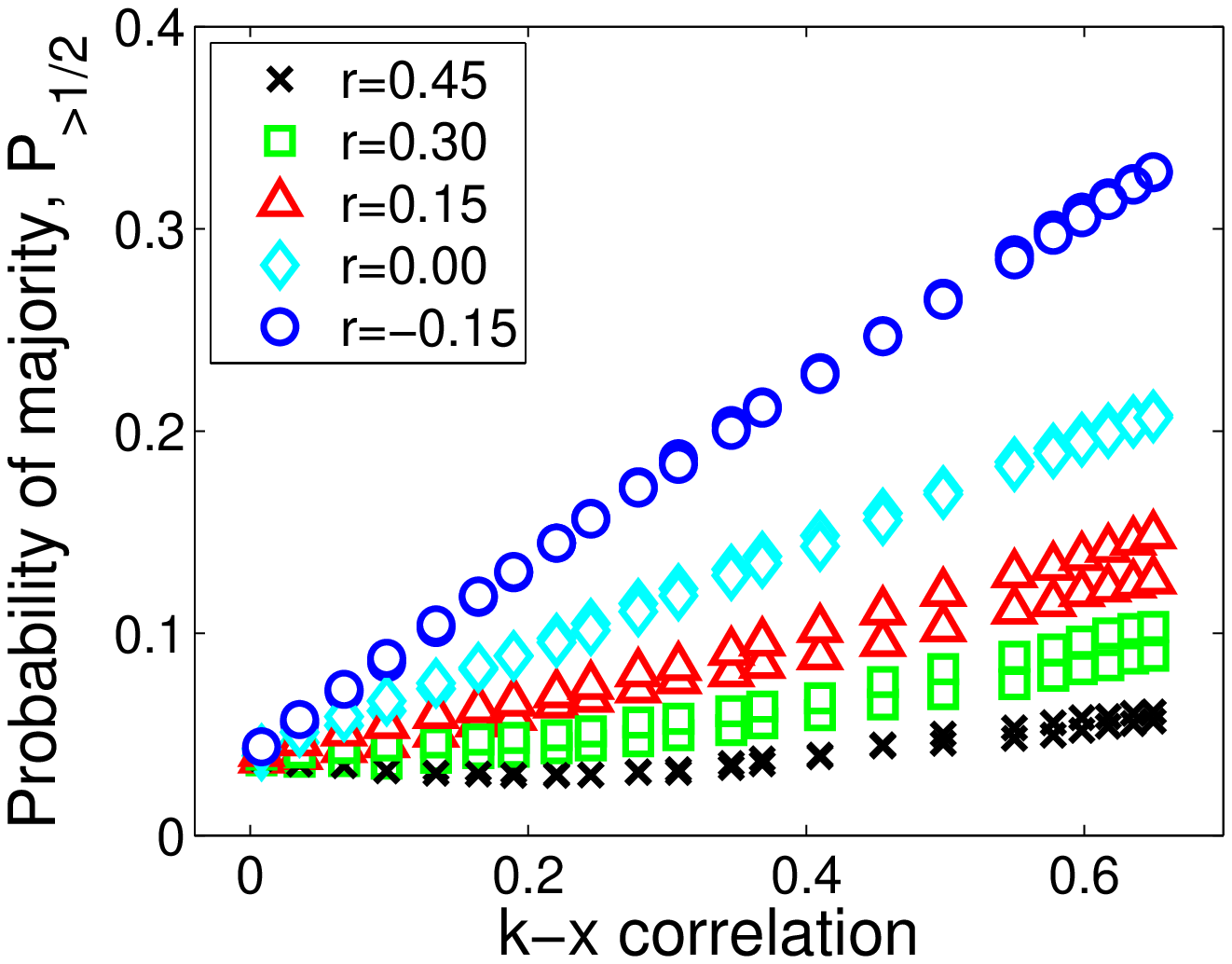}}
\end{tabular}
\caption{Strength of the majority illusion in synthetic scale-free networks showing that networks with the same degree sequence and assortativity can manifest differing levels of the paradox. Identical symbols in the same plot are for networks with the same assortativity. }
    \label{fig:idseq}
\end{figure*}

A network is not fully specified by its degree sequence and degree assortativity $r$. In fact, many different structures are possible with an identical degree sequence and $r$. These structural difference may affect the ``majority illusion.'' Here, we report some comparison of selected degree sequences in scale-free synthetic networks.

We generated scale-free networks with some degree sequence and assortativity. We then used edge rewiring to change network's structure while keeping the assortativity $r$ (and degree sequence) the same. Existing structural constraints restrict the range of assortativity values that a given degree sequence may attain. Thus there are fewer choices for extreme values of assortativity. Figure~\ref{fig:idseq} reports the fraction of nodes in the paradox regime in these networks. Identical symbols are for the same value of assortativity. Results don't change much in cases where structural constraints prevent varying the structure while keeping assortativity fixed. On the other hand, the fraction of nodes in the paradox regime can vary somewhat in the mid-assortativity range.


\end{document}